\documentclass[preprint,authoryear,12pt]{elsarticle}

\usepackage{amssymb}

\usepackage{amsmath,amscd}
\usepackage{empheq,array}
\usepackage{mathrsfs,stmaryrd}
\usepackage{relsize,exscale}
\usepackage{theorem}

\usepackage{
   nicefrac, 
   upgreek,
}
\usepackage[loose,nice]{units}
\usepackage{graphicx,psfrag}

\DeclareMathOperator{\Sp}{tr}

\DeclareMathOperator{\x}{\mathbf{x}}

\DeclareMathOperator{\diag}{diag}

\theoremstyle{plain}
\newtheorem{remark}{Remark}

\newbox\JIGAMMa
\newbox\JIGAMMb
\newbox\JIGAMMc
\newbox\JIGAMMd
\newbox\JIGAMMz
\newbox\LLbox
\newbox\LLboxh
\newbox\SLhilfbox
\newbox\SLubox
\newbox\SLobox
\newbox\SLergebnis
\newbox\TENbox
\newif\ifSLoben
\newif\ifSLunten
\newdimen\JIGAMMdimen
\newdimen\JIhsize\relax\JIhsize=\hsize
\newdimen\SLrandausgleich
\newdimen\SLhoehe
\newdimen\SLeffbreite
\newdimen\SLuvorschub
\newdimen\SLmvorschub
\newdimen\SLovorschub
\newdimen\SLsp
\def\pkt{\cdot}
\def\ppkt{\mathbin{\mathord{\cdot}\mathord{\cdot}}}
\setbox\JIGAMMa = \hbox{$\scriptscriptstyle c$} \setbox\JIGAMMz =
\hbox{\hskip-.35pt\vrule width .25pt\hskip-.35pt
                        \vbox to1.2\ht\JIGAMMa{\vskip-.125pt
                             \hrule width1.2\ht\JIGAMMa height.25pt
                             \vfill
                             \hrule width1.2\ht\JIGAMMa height.25pt
                             \vskip-.125pt}%
                        \hskip-.125pt\vrule width .25pt\hskip-.125pt}
\def\Oldroy#1#2#3{\STAPEL{#1}!_\SLstrich!_\SLstrich!^\circ{}
                  \ifx #2,{}_{\copy\JIGAMMz}%
                  \else \mskip1mu{}^{\copy\JIGAMMz}\fi
                  \mskip1mu\ifx #3,{}_{\copy\JIGAMMz}%
                           \else {}^{\copy\JIGAMMz}\fi }
\def\OP#1#2{\ifnum#1=1{\rm S}
            \else\ifnum#1=2{\rm S}^\JIv
                 \else\ifnum#1=3{\rm S}^T
                      \else{{\rm S}^T}^\JIv
            \fi\fi\fi\LL{{#2}}\RR}
\edef\JIminus{{\setbox\JIGAMMa=\hbox{$\scriptstyle x$}%
           \hbox{\hskip .10\wd\JIGAMMa
                 \vbox{\hrule width .6\wd\JIGAMMa height .07\wd\JIGAMMa
                       \vskip.53\ht\JIGAMMa}%
                 \hskip .10\wd\JIGAMMa}}}
\edef\JIv{{\JIminus 1}}

\def\LL#1\RR{\setbox\LLbox =\hbox{\mathsurround=0pt$\displaystyle
                                              \left(#1\right)$}%
       \setbox\LLboxh=\hbox{\mathsurround=0pt%
                  $\displaystyle{\left(%
                      \vrule width 0pt height\ht\LLbox depth\dp\LLbox
                      \right)}$}%
       \left(\hskip-.3\wd\LLboxh\relax\copy\LLbox
              \hskip-.3\wd\LLboxh\relax\right)}
\def\ZBOX#1#2#3{\def#3{}%
                \setbox#1 = #2
                \def#3{ to \wd#1}%
                \setbox#1 = #2}
\def\SLdreieck{\setbox\TENbox=\hbox{\fontscsy\char\dq 34}%
                 \dp\TENbox = 0pt
                 \hbox{\hskip -2\SLrandausgleich
                       \box\TENbox
                       \hskip -2\SLrandausgleich}}
\def\SLtilde{\setbox\TENbox=\hbox{\fontscex\char\dq 65}%
                    \vbox{\vskip-.03\ht\TENbox
                          \hbox{\hskip -1\SLrandausgleich
                                \copy\TENbox
                                \hskip -1\SLrandausgleich}
                          \vskip -.86\ht\TENbox}}
\def\SLstrich{\vrule width \SLeffbreite height.4pt}
\def\SLpunkt{{\vbox{\hbox{$\displaystyle.$}\vskip.03cm}}}
\def\SLabstand{\vskip .404pt}
\def\SLzwischen{\vskip 1.372pt}
\font\fontscsy=cmsy6 \font\fontscex=cmex10 scaled 1200
\def\STAPEL#1{\def\SLkern{#1}%
              \futurelet\next\SLpruef
               A_0   _0    :B_0   _-.17 :C_.05 _-.15 :D_0   _-.2
              :E_0   _-.2  :F_0   _-.21 :G_0   _-.15 :H_0   _-.23
              :I_.2  _.15  :J_.05 _-.1  :K_0   _-.22 :L_0   _-.1
              :M_0   _-.23 :N_0   _-.25 :O_.05 _-.2  :P_0   _-.21
              :Q_.05 _-.2  :R_0   _-.03 :S_0   _-.15 :T_.2  _0
              :U_.1  _-.1  :V_.1  _-.15 :W_.1  _-.2  :X_0   _-.22
              :Y_.16 _-.15 :Z_0   _-.25
              :a_.05 _-.05 :b_.05 _0    :c_.05 _.05  :d_0   _-.05
              :e_.07 _0    :f_0   _-.15 :g_.04 _-.2  :h_0   _-.07
              :i_.05 _0    :j_.08 _-.1  :k_0   _-.1  :l_.2  _.15
              :m_0   _-.1  :n_0   _-.1  :o_0   _-.1  :p_.15 _0
              :q_.1  _0    :r_.1  _-.1  :s_0   _-.2  :t_.1  _.05
              :u_0   _-.1  :v_0   _-.2  :w_0   _-.2  :x_.04 _-.14
              :y_.15 _-.05 :z_0   _-.15
              :\mit\Phi_.08 _-.1    :\mit\Omega_0 _-.2   :\varXi_.00 _-.2
              :\alpha_0 _-.2        :\gamma_.1 _-.1      :\varepsilon_.1 _-.1
              :\epsilon_.05 _-.05   :\eta_.05 _-.15      :\lambda_0 _0
              :\mu_0 _-.25          :\nu_0 _-.2          :\varSigma_-.03 _-.2
              :\varrho_.00 _-.2     :\sigma_.1 _-.2      :\tau_.15 _-.15
              :\varphi_.2 _-.1      :\omega_.1 _-.1      :\mit\Gamma_-.1 _-.1
              :\Lambda_0 _0         :\Gam_0 _0           :\Lam_0 _0
              :\SLsuchende
              \def\SLtrick{\noexpand\SLtrick\noexpand}%
                \def\SLdummy{\noexpand\SLdummy}%
                \edef\SLoboxinhalt{}\edef\SLuboxinhalt{}%
                \SLobenfalse\SLuntenfalse
                \futurelet\next\SLsuchruf}
  \def\SLsuchruf{\ifx\next !\let\next\SLexpand
                 \else\let\next\SLerzeug\fi\next}
  \def\SLexpand#1#2#3{\ifx #2\sb\ifSLunten\let\SLspeicher\SLuboxinhalt
                   \else\def\SLspeicher{\SLtrick\SLabstand}\fi
                   \edef\SLuboxinhalt{%
                       \SLspeicher
                       \SLtrick\SLzwischen
                       \hbox\SLdummy{\hfil\mathsurround=0pt
$\SLtrick\scriptstyle\SLtrick#3$%
                                     \hfil}}%
                   \SLuntentrue%
                      \else\ifSLoben\let\SLspeicher\SLoboxinhalt
                   \else\def\SLspeicher{\SLtrick\SLabstand}\fi
                   \edef\SLoboxinhalt{%
                       \hbox\SLdummy{\hfil\mathsurround=0pt
$\SLtrick\scriptstyle\SLtrick#3$%
                                     \hfil}%
                       \SLtrick\SLzwischen
                       \SLspeicher}%
                   \SLobentrue\fi\futurelet\next\SLsuchruf}
  \def\SLerzeug{\def\SLtrick{}
                \setbox\SLhilfbox=\hbox{$\displaystyle{E}$}%
                \SLrandausgleich=.04\wd\SLhilfbox
                      \setbox\SLhilfbox=%
                         \hbox{\hskip -1\SLrandausgleich
                          \mathsurround=0pt$\displaystyle{\SLkern}$%
                               \hskip -1\SLrandausgleich}%
                      \SLhoehe = \ht\SLhilfbox
                      \advance\SLhoehe by \dp\SLhilfbox
                      \SLeffbreite = \wd\SLhilfbox
                      \advance\SLeffbreite by \SLab\SLhoehe
                      \ZBOX\SLubox{\vbox{\offinterlineskip
                                         \SLuboxinhalt
                                         \hrule height 0pt}}\SLdummy
                      \ZBOX\SLobox{\vbox{\offinterlineskip
                                         \SLoboxinhalt
                                         \hrule height 0pt}}\SLdummy
                      \SLsp = \SLzu\SLhoehe
                      \advance\SLsp by -.5\SLeffbreite
                      \SLuvorschub = -1\SLsp
                      \advance\SLuvorschub by -.5\wd\SLubox
                      \SLovorschub = -1\SLsp
                      \advance\SLovorschub by -.5\wd\SLobox
                      \advance\SLovorschub by .26\SLhoehe
                      \ifdim\SLuvorschub > \SLovorschub
                         \SLsp = \SLovorschub
                      \else
                         \SLsp = \SLuvorschub
                      \fi
                      \ifdim\SLsp < 0pt%
                         \advance\SLuvorschub by -1\SLsp
                         \SLmvorschub = -1\SLsp
                         \advance\SLovorschub by -1\SLsp
                      \else
                         \SLmvorschub = 0pt
                      \fi
                      \setbox\SLergebnis = \hbox{%
                         \offinterlineskip
                         \hskip\SLrandausgleich\relax
                         \vbox to 0pt{%
                            \vskip -1\ht\SLobox
                            \vskip -1\ht\SLhilfbox
\hbox{\hskip\SLovorschub\copy\SLobox\hfil}%
                            \hbox{\hskip\SLmvorschub\copy\SLhilfbox
                                  \hfil}%
\hbox{\hskip\SLuvorschub\copy\SLubox\hfil}%
                            \vss}%
                         \hskip\SLrandausgleich}%
                      \SLsp = \dp\SLhilfbox
                      \advance\SLsp by \ht\SLubox
                      \dp\SLergebnis = \SLsp
                      \SLsp = \ht\SLhilfbox
                      \advance\SLsp by \ht\SLobox
                      \ht\SLergebnis = \SLsp
                      \box\SLergebnis{}}
  \def\SLpruef{\ifx\next\SLsuchende\def\SLzu{0}\def\SLab{0}%
                  \def\next##1\SLsuchende{\relax}%
               \else\let\next\SLvergl
               \fi\next}
  \def\SLvergl#1_#2_#3:{\def\SLv{#1}%
                        \ifx\SLkern\SLv\def\SLzu{#2}\def\SLab{#3}%
\def\next##1\SLsuchende{\relax}%
                        \else\def\next{\futurelet\next\SLpruef}
                        \fi\next}
\def\PKT#1{#1!^\SLpunkt}

\newbox\minusbox
\def\minus{\mathchoice{\minusarb\displaystyle}%
                      {\minusarb\textstyle}%
                      {\minusarb\scriptstyle}%
                      {\minusarb\scriptscriptstyle}}
  \def\minusarb#1{\setbox\minusbox=\hbox{$#1x$}%
                  \hbox{\hskip .10\wd\minusbox
                        \vbox{\hrule width .6\wd\minusbox
                                     height .07\wd\minusbox
                              \vskip.53\ht\minusbox}%
                        \hskip .10\wd\minusbox}}

\def\Vek#1{\STAPEL {#1}!_\SLstrich}
\def\Ten2#1{\STAPEL {#1}!_\SLstrich!_\SLstrich}

\def\e{\STAPEL e!_\SLstrich!_\SLstrich}

\def\D{\STAPEL D!_\SLstrich!_\SLstrich}

\newcommand{\inv}{{\minus 1}}
\newcommand{\T}{\text{T}}                                                                 
\renewcommand{\D}{\text{D}}                                                               
\newcommand{\w}{\text{w}}                                                                 
\renewcommand{\c}{\text{c}}                                                               
\renewcommand{\t}{\text{t}}                                                               
\newcommand{\mfrac}[2]{\mathlarger{\tfrac{#1}{#2}}}                                       
\newcommand{\doverd}[2]{\dfrac{{\rm d}#1}{{\rm d}#2}}

\newcommand{\kB}{k_{\text{B}}}
\newcommand{\Cut}{\mathscr{C}} 
\newcommand{\Resp}{\mathscr{R}} 
\newcommand{\Inten}{\mathscr{I}}
\newcommand{\MPa}{\unit{MPa}}
\newcommand{\Paragraph}[1]{\vspace{2mm}\emph{\underline{#1}}\par\vspace{2mm}}

\newcommand{\Eng}{\negthickspace\negthickspace}                                           
\newcommand{\en}{\hspace{-0.8mm}}                                                         
\newcommand{\En}{\hspace{-1.2mm}}                                                         

\newcommand{\Matr}[1]{\mathbf{#1}}                                                        

\renewcommand{\i}{\text{i}}                                 
\renewcommand{\d}{\text{d}}                                 
\renewcommand{\e}{\text{e}}                                 
\newcommand{\ie}{\text{ie}}                                 
\newcommand{\ii}{\text{ii}}                                 
\renewcommand{\t}{\text{t}}                                 

\newcommand{\vCie}{\check{\Ten2C\,}{_{\hspace{-0.7mm}\ie}}}

\newcommand{\CII}{\Ten2C{_{\hspace{-0.3mm}\ii}}}

\newcommand{\vDii}{\check{\Ten2D\,}{_{\hspace{-0.8mm}\ii}}}

\newcommand{\vLii}{\check{\Ten2L\,}{_{\hspace{-0.6mm}\ii}}}
\newcommand{\Ci}{\Ten2C{_{\hspace{-0.3mm}\i}}}

\newcommand{\Di}{\hat{\Ten2D\,}{_{\hspace{-0.8mm}\i}}}

\newcommand{\Gam}{\Ten2\varGamma!^{\boldsymbol{\wedge}}}

\newcommand{\Ce}{\hat{\Ten2C\, }{_{\hspace{-0.7mm}\e}}}

\newcommand{\Li}{\hat{\Ten2L\,}{_{\hspace{-0.3mm}\i}}}

\newcommand{\Fk}[1]{{\Ten2F}{_{\hspace{-0.5mm}#1}}}

\newcommand{\FKK}[1]{{\Ten2F}{_{\hspace{-0.4mm}#1}}}
\newcommand{\DotFk}[1]{\PKT{\Ten2F}{_{\hspace{-0.4mm}#1}}}

\newcommand{\DotFKK}[1]{\PKT{\Ten2F}{_{\hspace{-0.4mm}#1}}}

\newcommand{\hS}{\hat{\Ten2S\,}\!}
\newcommand{\PI}{\hat{\Ten2\varSigma\,}\!}

\newcommand{\hX}{\hat{\Ten2X\,}\!}

\newcommand{\vX}{\check{\Ten2X\, }\!}

\newcommand{\vP}{\check{\Ten2\varXi\, }\!}

\newcommand{\psiel}{\psi_{\text{el}}}
\newcommand{\psiiso}{\psi_{\text{iso}}}
\newcommand{\psikin}[1]{\overset{\Eng\scriptscriptstyle{\mathrm{#1}}}{\psi_{\text{kin}}}}

\newcommand{\kreiss}[1]
         {\unitlength1ex 
       \begin{picture}(0.9,0.9)           
       \put(0.45,0.45){\circle{0.9}}
       \put(0.45,0.45){\makebox(0,0){#1}}
       \end{picture}}

\newcount\OLordnung
\def\Oldroyd#1{%
   \unsichtbartrue\def\PGstrichdicke{.25pt}
   \OLordnung=0
   \def\OLarg{#1}
   \def\OLbef{}
   \def\OLu{\noexpand\OLu}
   \def\OLplatz{\noexpand\OLplatz}
   \def\OLo{\noexpand\OLo}
   \futurelet\OLnaechster\OLpruef}
\def\OLpruef{%
   \ifx\OLnaechster,
      \advance\OLordnung by 1
      \ifnum\OLordnung=1
         \edef\OLbef{\OLbef\OLu}%
      \else
         \edef\OLbef{\OLbef\OLplatz\OLu}%
      \fi
      \let\OLnext\OLpruefweiter
   \else
      \ifx\OLnaechster'
         \advance\OLordnung by 1
         \edef\OLbef{\OLbef\OLplatz\OLo}%
         \let\OLnext\OLpruefweiter
      \else
         \def\OLu{{}_{\Paragraph\Quad}}%
         \def\OLo{{}^{\Paragraph\Quad}}%
         \def\OLplatz{\mskip1mu}%
         \def\OLnext{%
            \expandafter\Ten\expandafter\OLordnung\OLarg%
                  !^{\kreiss{}}
            \OLbef}%
      \fi
   \fi
   \OLnext}
\def\OLpruefweiter#1{\futurelet\OLnaechster\OLpruef}

\newcommand{\kreis}[1]                                                                                                     
         {\unitlength1ex                                                                                                         
       \begin{picture}(1.0,1.0)
       \put(0.5,0.5){\circle{1.0}}
       \put(0.5,0.5){\makebox(0,0){#1}}
       \end{picture}}

\def\silb#1{\kreis{$\scriptscriptstyle{\text{#1}}$}}

\newcount\OLordnung
\def\OldroydZ#1#2-{%
   \unsichtbartrue\def\PGstrichdicke{.25pt}
   \OLordnung=0
   \def\OLargA{#1}
   \def\OLargB{#2}
   \def\OLbefZ{}
   \def\OLu{\noexpand\OLu}
   \def\OLplatz{\noexpand\OLplatz}
   \def\OLo{\noexpand\OLo}
   \futurelet\OLnaechster\OLpruefZ}
\def\OLpruefZ{%
   \ifx\OLnaechster,
      \advance\OLordnung by 1
      \ifnum\OLordnung=1
         \edef\OLbefZ{\OLbefZ\OLu}%
      \else
         \edef\OLbefZ{\OLbefZ\OLplatz\OLu}%
      \fi
      \let\OLnext\OLpruefweiter
   \else
      \ifx\OLnaechster'
         \advance\OLordnung by 1
         \edef\OLbefZ{\OLbefZ\OLplatz\OLo}%
         \let\OLnext\OLpruefweiter
      \else
         \def\OLu{{}_{\Paragraph\Quad}}%
         \def\OLo{{}^{\Paragraph\Quad}}%
         \def\OLplatz{\mskip1mu}%
         \def\OLnext{%
            \expandafter\Ten\expandafter\OLordnung\OLargA%
                  !^{\silb {\OLargB}}
            \OLbefZ}%
      \fi
   \fi
   \OLnext}
\def\OLpruefweiter#1{\futurelet\OLnaechster\OLpruefZ}

\journal{}

\begin{document}

\begin{frontmatter}



\title{On a dislocation density based two-phase plasticity model: refinement and extension to non-proportional loading}


\author[adr]{C.B. Silbermann\corref{cor1}}
\cortext[cor1]{Tel.: +49-371-531-38845; Fax: +49-371-531-838845}
\makeatletter
\ead{christian.silbermann@mb.tu-chemnitz.de}
\makeatother
\author[adr]{A.V. Shutov}
\author[adr]{J. Ihlemann}
\address[adr]{Chemnitz University of Technology, Institute of Mechanics and Thermodynamics, Professorship of Solid Mechanics, Strasse der Nationen 62, 09111 Chemnitz, Germany}

\begin{abstract}

The two-phase composite approach of \citet{Estrin1998} describes an evolving dislocation cell structure. \citet{Mckenzie2007} enhanced the model to capture the effects of hydrostatic pressure and temperature during severe plastic deformation. The goal of the present study is to incorporate this microstructural model into the macroscopic viscoplasticity framework proposed by \citet{Shutov2008a}. Thereby, the two-phase composite approach is examined carefully. Both physical and numerical drawbacks are revealed and possible solutions are presented, thus leading to a refined micro model. Moreover, some improvements concerning reliable parameter identification are suggested as well. The material parameters of the refined micro model are identified for an aluminum alloy using TEM cell size measurements. Then, an extension to non-proportional deformation is performed in such a way that the evolution of dislocation densities becomes sensitive to load path changes. Experimental findings suggest that such deformation modes can significantly influence the evolution of microstructure, including the dissolution of cells and the reduction of total dislocation density shortly after the load path change. In order to capture such effects, some tensor-valued state variables are introduced which couple the refined micro model with the macroscopic viscoplasticity model. As a result, a new system of constitutive equations is obtained. In order to demonstrate its capability to respond to load path changes, load cases as typical for Equal Channel Angular Pressing (ECAP) are considered. The obtained evolution of dislocation populations differs signficantly depending on which ECAP route is applied.

\end{abstract}

\begin{keyword}
Two-phase composite model, dislocation cells, viscoplasticity, non-proportional loading, ECAP
\end{keyword}

\end{frontmatter}


\section{Introduction}

An adequate description of evolving dislocation structures is of great importance for physically based modeling of many metallic materials. There are sophisticated multi-scale approaches to model dislocation dynamics in a discrete or continuous manner \citep[e.g.][]{Devincre2001, Groh2009b, Cai2006, Le2008}. Thus, precious insights into the collective behaviour of dislocations can be obtained \citep[for an overwiew, see][]{Bulatov2006}. However, the scale bridging to engineering plasticity applications where large deformations and long time frames occur is still an open problem.\par
The number of degrees of freedom can be drastically decreased by introducing dislocation densities as state variables\footnote{Frequently, they are also referred to as inner or structural variables.} into continuum scale models. Corresponding evolution equations can be deduced from microstructural considerations. Depending on the number of introduced dislocation populations these approaches are called one-parameter-model (only total dislocation density) or multi-parameter-models, respectivly \citep[cf.][]{Goerdeler2001, Ding2001}. Within the last decade, this approach has been used by many researchers to include an evolving dislocation structure in continuum-scale modeling of metal plasticity. In particular, such information can be incorporated within models of crystal plasticity \citep[see][and references therein]{Gerard2013} or even purely phenomenological modeling \citep[see][]{Shutov2012}.\par
There are numerous theoretical and experimental indications that dislocations within crystals constitute self-organizing systems \citep{Amodeo1990a,Yamaki2006,Gregor1998}. As a characteristic feature, they tend to distribute inhomogeneously and thus form spatial patterns. Among others, the three-dimensional cell structure \citep{Langlois2003} is typical for a class of metals and alloys. After the dislocation structure has emerged, further monotonic deformation results in the shrinkage of the structural characteristics: dislocation spacing, wall thickness and cell size. In the case of non-monotonic deformation the behavior seems to be different. Experimental findings \citep{Hasegawa1975,Hasegawa1986} suggest that load path changes can lead to the dissolution or disruption of cells.\par 
The evolving cell structure, i.e. the formation, sharpening and possible dissolution of the cells, governs the strength behaviour on the macroscopic level. Additionally, dislocation cells can serve as precursor of forthcoming subgrains even leading to nanocrystalline microstructures under severe plastic deformations \citep{Toth2010}. In particular, Equal Channel Angular Pressing (ECAP) is a promising technique which allows development of an ultrafine-grained microstructure without net change in the billet's shape \citep{Segal1995}.\par
The present study is organized as follows: the two-phase composite approach of \citet{Estrin1998} is outlined and carefully analyzed (cf. Section 2). Both physical and numerical drawbacks are identified and possible solutions are presented, thus leading to a refined model. The problem of a reliable parameter identification is discussed in Section 3. There, material parameters of the refined model are identified for the aluminum alloy AA 6016. In Section 4 the model is extended to non-proportional deformation in such a way that the evolution of dislocation densities becomes sensitive to load path changes. Towards that end, the refined micro model is coupled with the macroscopic viscoplasticity model of \citet{Shutov2008a}, which is also outlined briefly. Finally, a new system of constitutive equations is obtained including the micro model of \citet{Mckenzie2007} as a special case. In Section 5, the evolution of dislocation populations during Equal Channel Angular Pressing (ECAP) is considered. In this context, the extension to non-proportional loading is especially important, since most commonly used ECAP routes involve distinct load path changes. Finally, simulation results for routes $A$, $C$, $B_c$, and $E$ are discussed.\par
Let us briefly introduce the notations used in this paper. A second rank tensor is denoted by a small or capital letter with two underscores, e.g. $\Ten2U$. The coefficients with respect to a certain Cartesian coordinate system $\Vek e_a$ are $U_{ab} = \Vek e_a\pkt\Ten2U\pkt\Vek e_b$. As usual, the trace of a tensor is denoted by the operator $\Sp$. This enables the definition of the double contraction of two second-rank tensors by $\Ten2U\ppkt\Ten2V=\Sp(\Ten2U\pkt\Ten2V)$. For the scalar product another double contraction in the form $\Ten2U:\Ten2V=\Sp(\Ten2U\!\pkt\!\Ten2V^\T)$ is used. Arranging second-rank tensor coefficients in a quadratic matrix is denoted by $[U_{ab}]$ or just $[U]$. Bold symbols indicate real-valued matrices regardless of where the coefficients arise from, e.g. $\Matr{x}=[1, 2, 3, 4]^\T$ is a 4-tuple or column matrix.

\section{Two-phase composite model}

The fundamental assumption of the dislocation density based two-phase composite model of \citet{Estrin1998} and \citet{Mckenzie2007} is that a dislocation cell structure has already formed. Among the variety of reported dislocation configurations \citep{Amodeo1990b}, cells are characterized by regions of lower dislocation density (the cell interior) surrounded by high dislocation density walls. The latter represent a topologically continuous skeleton \citep{Mckenzie2007}. Hence, the polycrystal can be thought of as a solid consisting of grains which again consist of dislocation cells. Because of the large difference in dislocation density, the cell wall and the interior show different mechanical behaviour. Thus, the solid can effectivly be treated as a two-phase composite \citep{Mughrabi1987}.
Whereas the effect of grain orientations can be incorporated by using crystal plasticity \citep[e.g.][]{Yalcinkaya2009}, further modeling is necessary to take into account the two phases due to dislocation cells.

\subsection{Outline of the existing model}

The ``unit cell'' of the dislocation arrangement is a single cube with an edge length of $d$ and a wall thickness of $w/2$ \footnote{In some studies, spherical dislocation cells are assumed in order to simplify the stress computation \citep{Sauzay2008,Brahme2011}.}. The cell wall volume $V_\w$ over the total volume $V_\t$ yields the volume fraction $f$ of cell walls:
\begin{align}\label{deff}
   f \coloneqq \frac{V_\w}{V_\t} = 1-\left(1-\frac{w}{d}\right)^3 \quad\Leftrightarrow\quad (d-w)^3=d^3(1-f)\ .
\end{align}
\begin{figure}[h]\centering
   \psfrag{A}[m][][1][0]{ $d$}
   \psfrag{B}[m][][1][0]{ $d$}
   \psfrag{C}[m][][1][0]{ $d$}
   \psfrag{D}[m][][1][0]{ $w/2$}
   \psfrag{E}[m][][1][0]{ $w$}
   \psfrag{F}[m][][1][0]{ interior}
   \psfrag{G}[m][][1][0]{ cell wall}
   \includegraphics[scale=0.6]{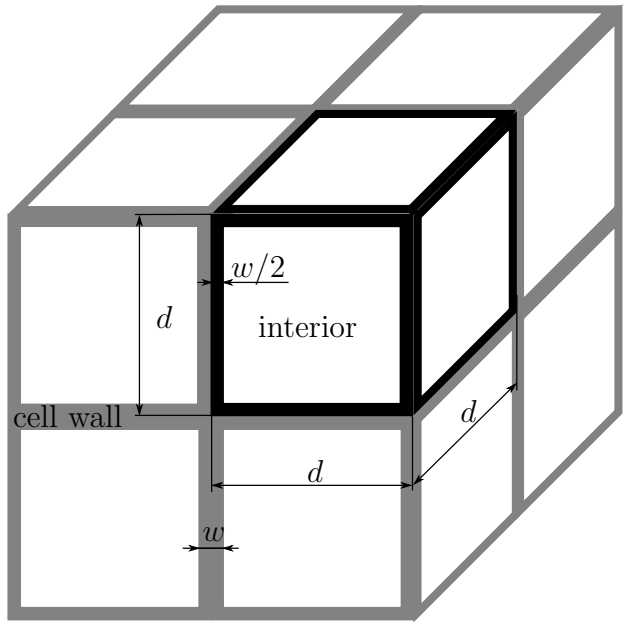}
   \caption[]{Scheme of the dislocation cell arrangement, the ``unit cell'' is colored black.}\label{3dZellen}
\end{figure}\noindent
As a modeling assumption, the dislocations are classified with respect to their position. A part $L_\w$ of the total dislocation length $L_\t$ is arranged within the cell wall volume $V_\w$ and the remaining part $L_\c$ is located in the cell interior $V_\c$. Hence, we can define the densities of two dislocation populations:
\begin{align}\label{disldensities}
   \rho_\w = \frac{L_\w}{V_\w} \quad,\quad \rho_\c = \frac{L_\c}{V_\c} \ .
\end{align}
Dividing the total dislocation length $L_\t=L_\w+L_\c$ by the total volume $V_\t= V_\w + V_\c$ yields the rule of mixtures for the total dislocation density:
\begin{equation}\label{rho_split}
   \rho_\t = f\rho_\w+(1-f)\rho_\c \ .
\end{equation}
By Eq. \eqref{rho_split} it is not explicitly determined how the populations behave, e.g. whether they are mobile or immobile\footnote{There exist alternative criteria to group dislocations, e.g. by their character (edge or screw) or mobility (mobile or immobile) \citep[e.g.][]{Amodeo1988,Austin2011,Austin2012,Gao2012}. Thus, one rather yields different \emph{species} (dislocations with the same behavior) instead of \emph{populations} (dislocations in the same domain). In contrast to Eq. \eqref{rho_split}, the total dislocation density is then just the sum of the densities of species and there is no rule of mixtures.}. However, \citet{Estrin1998} considered the dislocations woven into the walls as immobilized. This point will be important for Section \ref{sec_FluxWC}.\par
\begin{figure}[h!]
\begin{minipage}{0.66\textwidth}
   \psfrag{A}[m][][1][0]{ $\rho_\w$}
   \psfrag{B}[m][][1][0]{ $\rho_\c$}
   \includegraphics[scale=0.28]{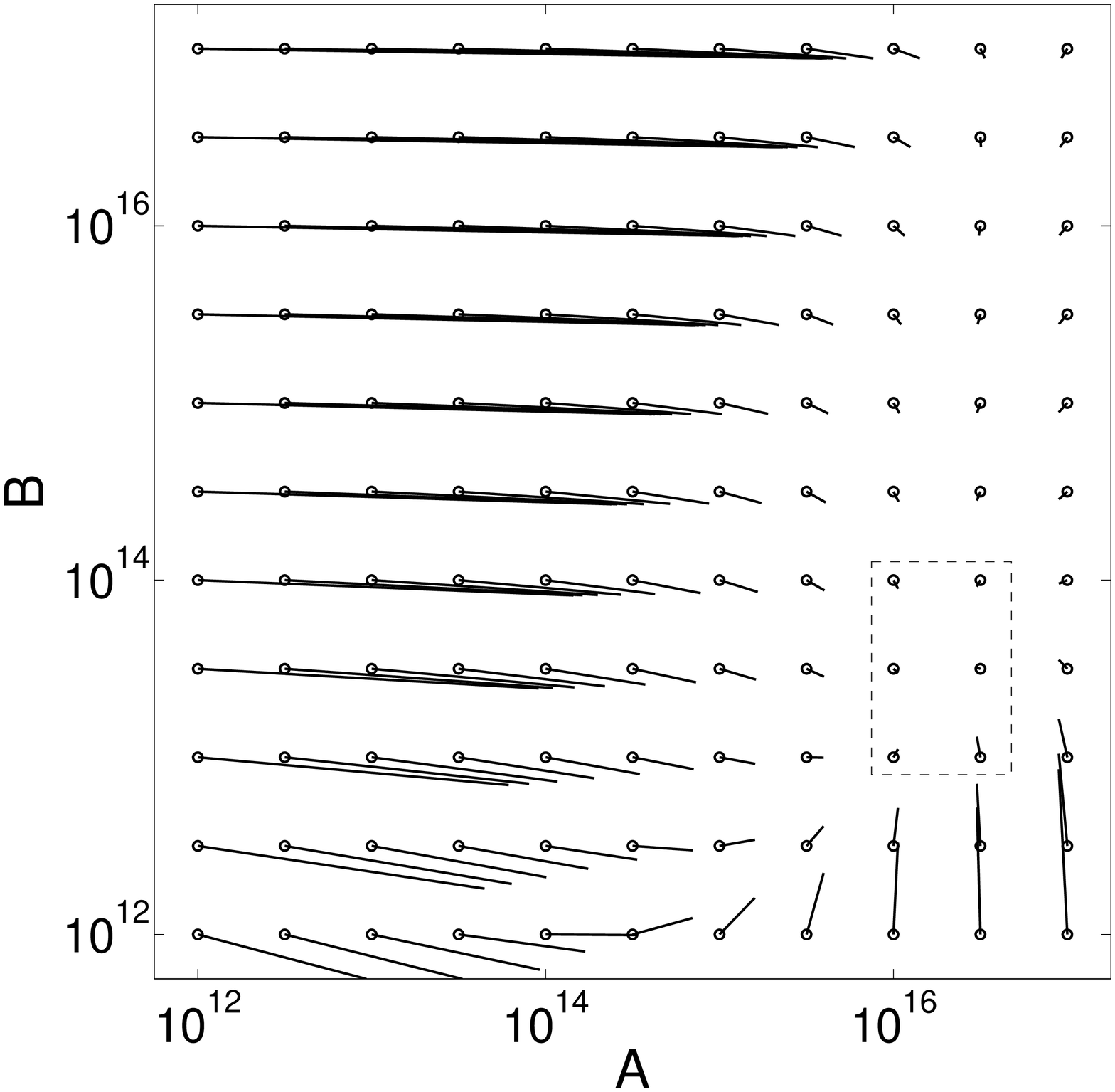}
\end{minipage}
\hspace{-15mm}
\begin{minipage}{0.33\textwidth}
   \psfrag{D}[m][][1][0]{ $\rho_\w$}
   \psfrag{C}[m][][1][0]{ $\rho_\c$}
   \includegraphics[scale=0.28]{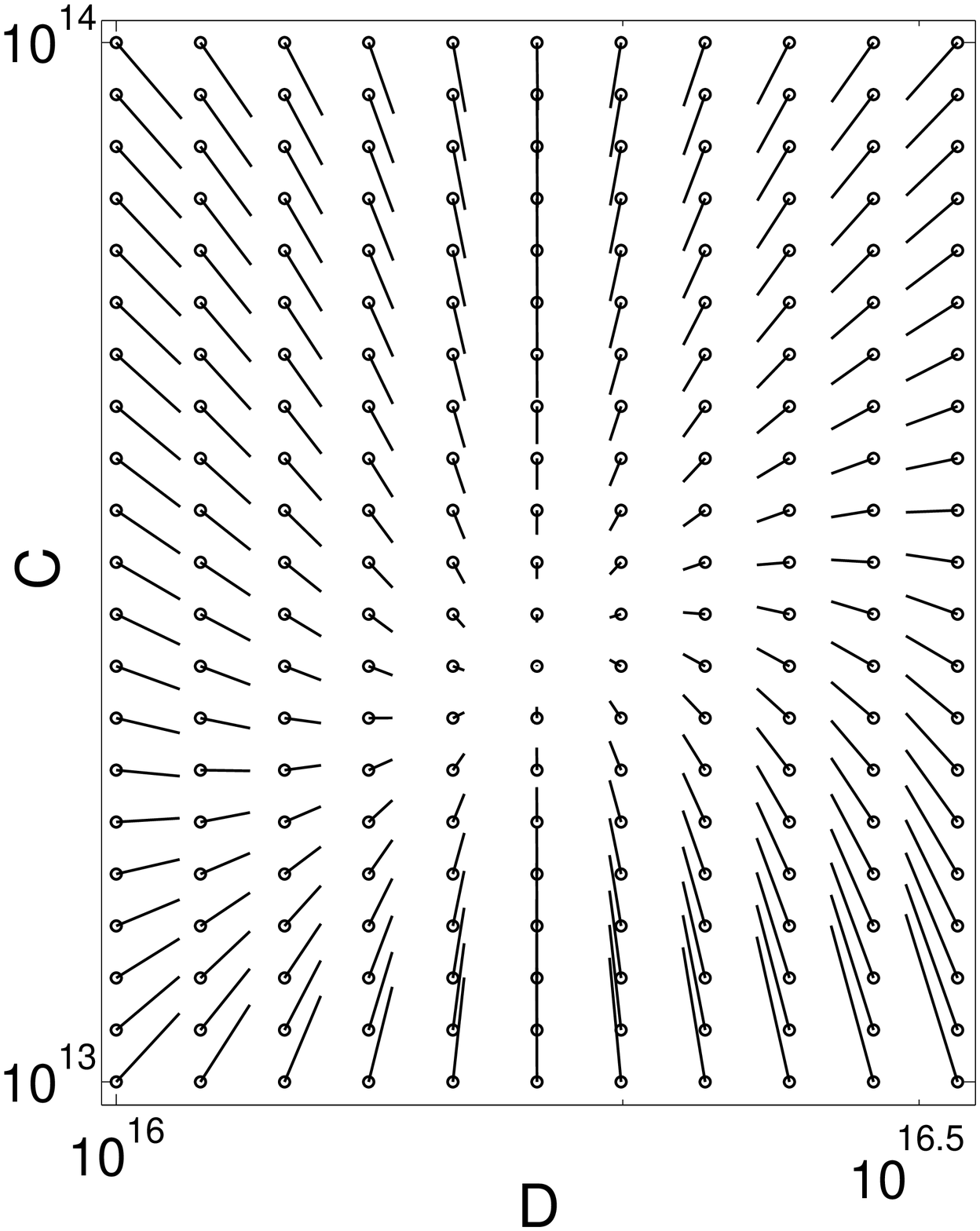}
\end{minipage}
\caption[]{Sketch of the velocity field in the phase space for $f=f_\infty=0.05$ (remaining parameters taken from Table \ref{constants}). The circles represent the grid. The zoom box on the right side reveals the fixed point. }\label{Velo_f05}
\end{figure}

By now, no evolution of cell size $d$ has been included explicitely. In the geometrical model (cf. Fig. \ref{3dZellen}) $w$ represents a fraction of the cell size $d$. Thus, changing the wall thickness $w$ or volume fraction $f$ does not affect the cell size. According to \citet{Amodeo1988} and \citet{Estrin1998}, it is assumed that the total dislocation density determines the cell size by the empirical relation
\begin{equation}
   d = K/\sqrt{\rho_\t} \ ,
\end{equation}
with a proportionality constant $K$. In order to derive evolution equations for the dislocation densities in the cell wall and the interior, elementary dislocation interactions are considered. Finally, one obtains the following system of evolution equations \citep[cf.][]{Estrin1998,Mckenzie2007}:
\begin{subequations}\label{EstrinOrig}
\begin{align}
   \doverd{\rho_\w}{t} =& + \frac{6\beta^\ast\dot{\gamma}_\c(1-f)^{2/3}}{bdf} + \frac{\sqrt{3}\beta^\ast\dot{\gamma}_\c(1-f)}{bf}\sqrt{\rho_\w}\notag\\
                        & -k_0\exp\left(- \frac{p\varOmega}{n_\w \kB T}\right) \left( \frac{\dot{\gamma}_\w}{\dot{\gamma}_0}\right)^{\inv /n_\w} \hspace{-1mm}\dot{\gamma}_\w \rho_\w\ , \label{Gl_rhow}\\
   \doverd{\rho_\c}{t} =& - \frac{6\beta^\ast\dot{\gamma}_\c}{bd(1-f)^{1/3}} + \frac{6\alpha^\ast\dot{\gamma}_\w}{\sqrt{3}b}\sqrt{\rho_\w} - k_0\left( \frac{\dot{\gamma}_\c}{\dot{\gamma}_0}\right)^{\inv /n_\c} \hspace{-1mm}\dot{\gamma}_\c\rho_\c  \label{Gl_rhoc}\ .
\end{align}\end{subequations}
The terms on the right hand  side represent the following processes: flux of dislocations from cell interior into the walls with $\beta^\ast$ representing the fraction of $L_\c$ woven into the wall (1\textsuperscript{st} term in \eqref{Gl_rhow} and \eqref{Gl_rhoc} respectivly); generation due to multiplication by {Frank-Read} sources (2\textsuperscript{nd} term in \eqref{Gl_rhow}, \eqref{Gl_rhoc}) with $\alpha^\ast$ as fraction of active sources; and annihilation supported by climb of edge dislocations (3\textsuperscript{rd} terms in \eqref{Gl_rhow}, \eqref{Gl_rhoc}). All these processes\footnote{Note that the \emph{nucleation} of new dislocations is not considered within this model.} are related to a change in $L_\w$ and $L_\c$ respectivly, which contributes to the change in $\rho_\w$ and $\rho_\c$. The shear rates in the cell wall and the interior are denoted by $\dot{\gamma}_\w$ and $\dot{\gamma}_\c$ respectively. The remaining parameters are explained in Table \ref{constants}.
\begin{table}[!h]
\centering
\caption{Parameters of the dislocation density based two-phase composite model from \citet{Estrin1998}}\label{constants}
\begin{tabular}{ | >{$}c<{$}| >{$}c<{$} | l | }
\firsthline
\text{Para.}               &  \text{Value/Range}                     & \text{Brief explanation} \\\hline\hline
   b                       &  2.86\!\cdot\!10^{-10}\unit{m}          & magnitude of {Burgers} vector for Al \\  
   \varOmega               &  1.2\!\cdot\!10^{-29}\unit{m}^3         & atomic volume for Al \\
   \kB                     &  1.38065\!\cdot\!10^{-23} \unitfrac{J}{K} & {Boltzmann}'s constant \\\hline
   \alpha                  &  >0                                     & controls the evolution of $f$ (cf. Eq. \eqref{fOrig}, \eqref{fevol}) \\
   f_0, f_\infty           &  (0,1)                                  & initial and final values for $f$ \\\hline
   p                       &  \geq0                                  & hydrostatic pressure, process dependent \\
   T                       &  >0                                     & absolute temperature, process dependent \\ 
   \dot{\gamma}            &  \geq0                                  & plastic shear strain rate, process dependent \\\hline
   \dot{\gamma}_0          &  >0                                     & reference shear strain rate \\
   \alpha^\ast             &  [0,1]                                  & fraction of active {Frank-Read} sources \\
   \beta^\ast              &  [0,1]                                  & emigrating fraction of $\rho_\c$ \\
   n_\c, n_\w              &  >0                                     & dynamic recovery exponents cell interior/wall\\
   k_0                     &  \geq0                                  & dynamic recovery constant \\
   K                       &  \geq0                                  & proportionality between $d$ and $1/\sqrt{\rho_\t}$ \\\hline
\end{tabular}
\end{table}
It can be argued that the ``stiff'' cell walls impose the same deformation on the cell interior \citep{Estrin1998} leading to the kinematic assumption $\dot{\gamma}_\w  = \dot{\gamma}_\c = \dot{\gamma}$. Thus, the evolution equations and the initial conditions can be written as
\begin{subequations}\label{EstrinOrig1}
\begin{align}
   \doverd{\rho_\w}{\gamma} =& + \frac{6\beta^\ast(1-f)^{2/3}}{bKf} \sqrt{\rho_\t} + \frac{\sqrt{3}\beta^\ast(1-f)}{bf}\sqrt{\rho_\w}\notag\\
                             &-k_0\exp\left(- \frac{p\varOmega}{n_\w \kB T}\right) \left( \frac{\dot{\gamma}}{\dot{\gamma}_0}\right)^{\inv /n_\w} \hspace{-1mm} \rho_\w\ , \label{Gl_rhow1}\\
   \doverd{\rho_\c}{\gamma} =& - \frac{6\beta^\ast}{bK(1-f)^{1/3}} \sqrt{\rho_\t} + \frac{6\alpha^\ast}{\sqrt{3}b}\sqrt{\rho_\w} - k_0\left( \frac{\dot{\gamma}}{\dot{\gamma}_0}\right)^{\inv /n_\c} \hspace{-1mm}\rho_\c \label{Gl_rhoc1} \\
                             &\rho_\w|_{\gamma=0} =\rho_{\w0} \ ,\ \rho_\c|_{\gamma=0}=\rho_{\c0} \ .
\end{align}\end{subequations}
Mathematically, \eqref{Gl_rhow1} and \eqref{Gl_rhoc1} constitute a set of two coupled, nonlinear, ordinary, first order differential equations for the variables $\rho_\w$ and $\rho_\c$. Since the wall volume fraction $f(t)$ is time-dependent, system \eqref{EstrinOrig1} is rheonome. Assuming the evolution of the cell wall volume fraction in the form \citep{Mckenzie2007}
\begin{equation}\label{fOrig}
   f = f_\infty + (f_0-f_\infty)\exp(-\alpha s)\ ,
\end{equation}
where $s$ denotes the accumulated plastic strain, initial value problem \eqref{EstrinOrig1} can be solved numerically. Consideration of the velocities $\left[{{\d\rho_\w}/{\d\gamma}},\ {{\d\rho_\c}/{\d\gamma}}\right]^\T$ in the phase space $[\rho_\w, \rho_\c]^\T$ for constant values of $f$ reveals a fixed point characteristic. In Fig. \ref{Velo_f05} the fixed point is characterized by a vanishing velocity. In particular, system \eqref{EstrinOrig1} exhibits fading memory behavior. Moreover, the solution is sensitive to the evolution of $f$ (cf. Fig. \ref{Velo_f05} and \ref{Velo_f07}).
\par

\subsection{Critical analysis, improvements and modifications}\label{subsec_Improve}

The model of \citet{Estrin1998} has been widely used in literature and adopted to many applications, especially in the field of severe plastic deformation \citep{Baik2003,Hosseini2009,Toth2010}. It has to be emphasized that the derivation of the evolution equations is based on clear physical considerations of few elementary dislocation interactions. Still, the model predictions are applicable over a wide range of strain and fit experimental results very well \citep{Mckenzie2007}.\par
Nevertheless, the model also has some weak points. It is known \citep{Madec2002} that cellular patterns form at the onset of stage III of the deformation of single crystals or at small plastic strains in multislip conditions. In any case, they do not emerge until the accumulated plastic strain reaches some value $s_{\text{crit}}>0$ \citep{Chinh2004}. Strictly speaking, the model is not valid before since it cannot take into account the precedent self-organization of dislocation structures. At the early stage of plastic deformation, alternative models should be implemented which account for the influence of the grain boundary topology on the dislocation storage \citep[cf.][]{Delannay2012}. Further, the characteristic microstructural quantities, such as the cell wall volume and the dislocation densities $\rho_\w$ and $\rho_\c$, are not clearly defined for $s<s_{\text{crit}}$ since they do not exist\footnote{Note that some authors suggest the initial condition $f=1$ in order to describe the state when no dislocation pattern is present \citep{Yalcinkaya2009}.}. Thus, it is unclear how to choose initial values for them. It is questionable whether measurements from the undeformed sample can be used. Fortunately, initial values are important only at the beginning due to the fading memory of the model. This aspect and a possible treatment are discussed in Section \ref{sec_Simul}.\par
\begin{figure}[h]\centering
   \psfrag{A}[m][][1][0]{ $\rho_\w$}
   \psfrag{B}[m][][1][0]{ $\rho_\c$}
   \includegraphics[scale=0.28]{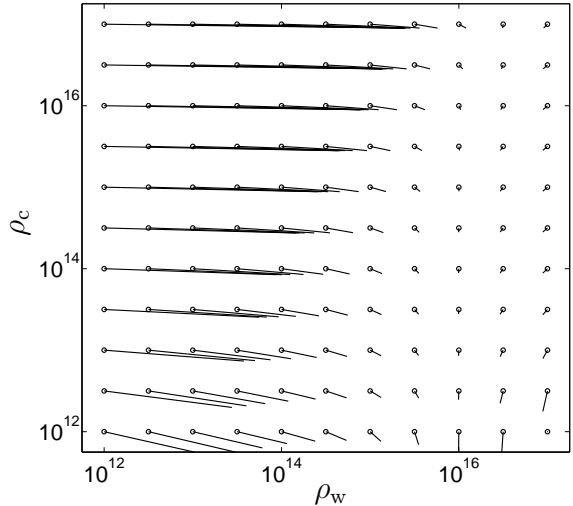}
   \caption[]{Sketch of the velocity field in the phase space for $f=f_0=0.07$ (remaining parameters taken from Table \ref{constants}).}\label{Velo_f07}
\end{figure}

Next, as seen in Fig. \ref{Velo_f05} and \ref{Velo_f07} the solution of system \eqref{EstrinOrig1} is very sensitive to the cell wall volume fraction $f$. Consequently, the evolution of $f$ should be measured accurately. However, this is related to huge experimental effort and not always justifiable in practise.\par
A minor but noticible problem stems from the treatment of the constant $\dot{\gamma}_0$. As it is merely used to produce a dimensionless term and set to $1/\unit{s}$ in \citep{Mckenzie2007} one should expect that the solution of system \eqref{EstrinOrig1} is independent from its choice. Because $\dot{\gamma}/\dot{\gamma}_0$ appears in \eqref{Gl_rhow1} and \eqref{Gl_rhoc1}, but with two different exponents $n_\w, n_\c$, the solution does depend on $\dot{\gamma}_0$. Indeed, given a set of parameters $k_0, n_\w, n_\c$ which correspond to a certain $\dot{\gamma}_0$, \emph{no equivalent} parameters can be found if $\dot{\gamma}_0$ is replaced by some new value. Hence, it should be considered as an additional parameter to be identified.\par
As an evident requirement, a physical model should produce reasonable results under any set of parameters within their range, but it was found that system \eqref{EstrinOrig1} can lead to \emph{negative values} of $\rho_\c$, e.g. using the parameters given from \citet{Mckenzie2007} and $\alpha=0.1$. This tendency is visible in Fig. \ref{Velo_f07} where the velocity vectors point downwards in the vicinity of $\rho_\c\rightarrow0$.
If a trajectory begins in these regions of phase space it is likely to cross $\rho_\c=0$ and go below. Obviously, the definition of dislocation density (cf. Eq. \eqref{disldensities}) requires it to be non-negative.\par
\begin{figure}[h!]\centering
   \psfrag{A}[m][][1][0]{ cell wall }
   \psfrag{B}[m][][1][0]{ interior }
   \psfrag{C}[m][][1][0]{ $d$}
   \psfrag{D}[m][][1][0]{ $v_\c\Delta t$}
   \psfrag{E}[m][][1][0]{ $\dfrac{w}{2}$}
   \includegraphics[scale=0.7]{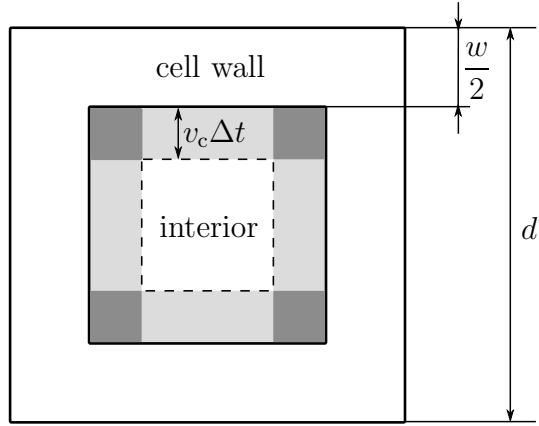}
   \caption[]{Cut of the ``elementary cell'' used to derive evolution equations \eqref{EstrinOrig1}. The shaded regions stand for the ``collecting area'' of dislocations which are woven into the cell walls. The dark grey regions can be neglected.}\label{fluxCW}
\end{figure}
\begin{figure}[h]\centering
   \psfrag{A}[m][][1][0]{ $\chi=b\rho_\c v_{\text{max}}/\dot{\gamma}_\c$ }
   \psfrag{B}[m][][1][0]{ $v_\c(\chi)/v_{\text{max}} $ }
   \psfrag{C}[l][][1][0]{ Orowan's equation}
   \psfrag{D}[l][][1][0]{ Modified equation}
   \includegraphics[scale=0.26]{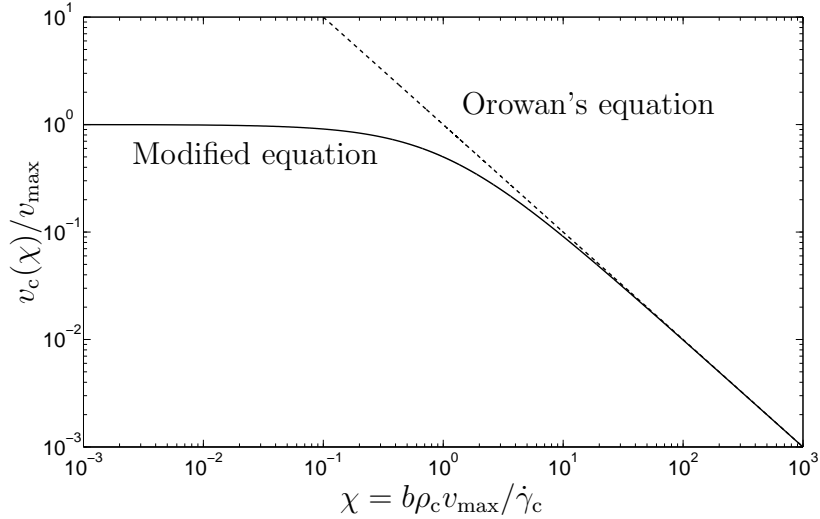}
   \caption[]{Logarithmic plot of {Orowan}'s equation (8) compared to the modified version given by (10).}\label{OrowanCurves}
\end{figure}

Therefore, one of the goals of the current publication is to modify the original evolution equations, such that the positiveness of dislocation densities will be guaranteed\footnote{In some applications, the initial values of dislocation densities may be chosen very close to zero \citep{Engels2012}. In this case there is a danger of obtaining negative dislocation densitites if the original model \eqref{EstrinOrig1} is used.}.
A closer look at the derivation from \citet{Estrin1998} reveals that the problem originates from the use of {Orowan}'s relation. More precisely, this equation is applied to relate the mean velocity $v_\c$ of \emph{mobile} dislocations in the cell interior with the macroscopic shear strain rate $\dot{\gamma}_\c$:
\begin{equation}\label{orowan}
   v_\c = \dot{\gamma}_\c/(b \rho_\c) \ .
\end{equation}
Next, this relation is used to derive the flux terms from cell interior to cell wall. Thereby, it is assumed that a fraction $\beta^\ast$ of all dislocations within a distance $v_\c\Delta t$ to the $6$ walls is woven into them, cf. Fig. \ref{fluxCW}. Here, $\Delta t$ is an infinitesimal amount of time. 
\noindent
That gives a ``collecting`` volume of $V_\c^\ast=6 v_\c\Delta t (d-w)^2$, neglecting the parts which are $(\Delta t)^2$ small (dark grey regions in Fig. \ref{fluxCW}). Hence, according to \citet{Estrin1998}, the loss of dislocation length in the cell interior is 
\begin{align}\label{term1}
   \Delta L_\c &= -\beta^\ast \rho_\c V_\c^\ast \overset{\eqref{orowan}}{=} -6\beta^\ast\rho_\c \left(\frac{\dot{\gamma}_\c}{\rho_\c b}\right)\Delta t (d-w)^2\ ,\notag\\
   \frac{\dot{L}_\c}{V_\c} &=-\frac{\Delta L_\c/ \Delta t}{(d-w)^3} \overset{\eqref{deff}}{=} -\frac{6\beta^\ast\dot{\gamma}_\c}{bd(1-f)^{1/3}} \ .
\end{align}
As can be seen, the flux $\dot{L}_\c/ V_\c$ is not vanishing as $\rho_\c\rightarrow0$. As a consequence, the loss of $\rho_\c$ can continue even when there are no longer any dislocations in the interior. Such behaviour cannot be expected for real physical systems. It is questionable whether {Orowan}'s equation is applicable in case of very low dislocation densities since according to \eqref{orowan}, $v_\c\rightarrow\infty$ as $\rho_\c\rightarrow0$. However, the dislocation velocity cannot exceed a certain limit due to drag forces. This gives rise to a modified {Orowan}-type equation where the maximum speed is bound to a value $v_{\text{max}}$. A simple approach used in the current study is given by
\begin{equation}
   \frac{1}{v_\c}  = \frac{1}{\dot{\gamma}_\c/(b\rho_\c)} + \frac{1}{v_{\text{max}}}\ .
\end{equation}
In other words, the travelling time of a mobile dislocation is assumed equal to the travelling time predicted by Orowan's relation plus a fixed minimum value. After a few manipulations, the dislocation velocity reads as:
\begin{align}\label{modorowan}
   v_\c & = \frac{\dot{\gamma}_\c}{b\rho_\c} \left[1+\frac{\dot{\gamma}_\c}{b\rho_\c v_{\text{max}}}\right]^{\inv} =\frac{\dot{\gamma}_\c}{b\rho_\c} \ \Cut(\rho_\c, \dot{\gamma}_\c)\ .
\end{align}
Here, $\Cut(\rho_\c, \dot{\gamma}_\c)=[1+{\dot{\gamma}_\c}/({b\rho_\c v_{\text{max}}})]^{\inv}$ stands for cut-off function since it ensures that $ v_\c\rightarrow v_{\text{max}}$ for $\rho_\c\rightarrow0$. Thus, the singularity is ``cut off''. Introducing this into the previous derivation of \eqref{term1} we get
\begin{equation}
   \frac{\dot{L}_\c}{V_\c}=-\frac{6\beta^\ast\dot{\gamma}_\c}{d(1-f)^{1/3}}\ \Cut(\rho_\c, \dot{\gamma}_\c)\ .
\end{equation}
The shear wave velocity represents a reasonable value for the upper bound of the mean dislocation velocity $v_{\text{max}}$ \citep[cf.][]{Johnston1959,Shehadeh2005}. 
From a mathematical point of view, any function $\Cut(\rho_\c, \dot{\gamma}_\c)$ with the property
\begin{equation}
   0 < \frac{\partial\Cut}{\partial\rho_\c}\bigg|_{\rho_\c=0} < \infty\label{Bedingung}
\end{equation}
could be chosen as an alternative to \eqref{modorowan}. However, equation \eqref{modorowan} offers the advantage of a clear interpretation. Using the non-dimensional number $\chi=b\rho_\c v_{\text{max}}/\dot{\gamma}_\c$ we find that for $\chi>99$ the difference between Eq. \eqref{modorowan} and {Orowan}'s equation \eqref{orowan} is less than $1\%$. Choosing $\chi=99$, $\dot{\gamma}_\c=0.1/\unit{s}$, $b$ from Table \ref{constants} and $v_{\text{max}}=3100\unitfrac{m}{s}$ we obtain $\rho_\c\approx10^{7}/\unit{m}^2$. Thus, assuming typical values for dislocation densities ranging from $10^{10}/\unit{m}^2$ in well-annealed crystals up to $10^{16}/\unit{m}^2$ for severely plastically deformed metals \citep{Hull2011}, Eq. \eqref{modorowan} is practically equivalent to {Orowan}'s equation (cf. Fig. \ref{OrowanCurves}). Another limiting case where the classical {Orowan} relation may not be applicable is given by high-rate plasticity with very large $\dot{\gamma}$ \citep{Krasnikov2011,Austin2012}.
\par\indent
Further, in \citep{Mckenzie2007} there is a discrepancy concerning the deformation variable $\gamma$ occuring in \eqref{Gl_rhow1} and \eqref{Gl_rhoc1}. The plastic shear rate is introduced in the model by using {Orowan}'s equation. It is not clear how to connect this quantity with an arbitrary macroscopic multiaxial state of deformation. In addition, it is not guaranteed that $\gamma$ increases monotonically with time. Consequently, the shear strain cannot be considered as a loading parameter, and the integration with respect to the time cannot be replaced by the integration with respect $\gamma$. This problem can be avoided by adopting the accumulated plastic strain $s$ instead. With the common multiplicative split of the deformation gradient $\Ten2F=\Fk{\e}\pkt\Fk{\i}$ we obtain the inelastic (plastic) strain rate tensor
\begin{equation}\label{DefD}
   \Di= \mfrac{1}{2}\left(\Li +\Li^\T \right) \quad\text{with}\quad \Li\coloneqq \DotFk{\i}\pkt{\Fk{\i}}^{\inv}
\end{equation}
as the symmetric part of the inelastic velocity gradient $\Li$. The accumulated plastic strain\footnote{Frequently, $s$ is also referred to as inelastic arc lentgh, true strain or deformation degree.} is then defined as
\begin{equation}\label{Defs}
   s(t)\coloneqq \int _{\tau=0}^{\tau=t}\dot{s}(\tau)\ d\tau \qquad\text{with}\qquad  \dot{s}(\tau)=\sqrt{\mfrac{2}{3}}\|\Di(\tau)\| \ ,
\end{equation}
where $\|\Ten2X\|$ denotes the Frobenius norm of a second-rank tensor $\Ten2X$, cf. definition \eqref{defFrob}. In fact, from the simulation results of \citet{Mckenzie2007} it is clear that the accumulated strain was used instead of the shear strain. Assuming plastic incompressibility and simple shear deformation in the $x-z$-plane offers a relationship between both quantities. With respect to a Cartesian coordinate system\footnote{Note that upright indices such as i are labels, wheras italic indices $a,b,\dotso$ stand for Cartesian coordinates $x,y,z$} one may consider
\begin{align}\label{Fform}
   \Big[F_{\i_{ab}}\Big] = 
   \begin{bmatrix*}[r]
    1 & 0 & \gamma(t) \\
    0 & 1   & 0 \\
    0 & 0   & 1 \\
\end{bmatrix*}\quad\overset{\eqref{DefD}}{\Rightarrow}\quad 
   \Big[\hat{D}_{\i_{ab}}\Big] =\frac{\dot{\gamma}}{2}
   \begin{bmatrix*}[c]
    0 & 0 & 1 \\
    0 & 0 & 0 \\
    1 & 0 & 0 \\
\end{bmatrix*}\ .
\end{align}
Taking the norm and evaluating \eqref{Defs} for the accumulated plastic strain, we obtain
\begin{equation}
   \dot{s} =|\dot{\gamma}| / \sqrt{3} \ .
\end{equation}
Thus, we replace $\dot{\gamma}$ with $\sqrt{3}\dot{s}$. In general, {Orowan}'s relation will be $\dot{s}=\xi\ v\rho b$ with the constant $\xi$ depending on the deformation mode. Here, $\xi=1/\sqrt{3}$ need not to be considered since it can be absorbed by model parameters. Consequently, the model exactly fulfills {Orowan}'s equation in the case of a plastic simple shear. In general, the kinematic relation $\dot{s}=v\rho b$ has to be considered as additional constitutive assumption.\par
The following considerations reveal another weakness of the model. Let us assume no production or annihilation of dislocations. In this case, there is just a dislocation flux stemming from the first terms in Eq. \eqref{Gl_rhow}, \eqref{Gl_rhoc}. Consequently, the total dislocation density (cf. Eq. \eqref{rho_split}) should then remain constant:
\begin{align}\label{rhot_zero}
   \dot{\rho_\t} & = \doverd{}{t}\big\{f\rho_\w+(1-f)\rho_\c\big\} \notag\\
                 & = \dot{f}(\rho_\w-\rho_\c) + f\dot{\rho}_\w + (1-f)\dot{\rho}_\c = 0 \ .
\end{align}
Substituting the flux terms from \eqref{EstrinOrig} into \eqref{rhot_zero} we get $\dot{f}(\rho_\w-\rho_\c)=0$ which is not fulfilled in general. This contradiction can be resolved by having a closer look at the rate of dislocation density. It is obtained by differentiating \eqref{disldensities} with respect to time. Though the total volume $V_\t$ is constant, the fraction $V_\w=fV_\t$ of the cell wall and $V_\c=(1-f)V_\t$ of the cell interior are time-dependent and we get
\begin{equation}\label{disldensrate}
   \dot{\rho}_\w = \frac{\dot{L}_\w}{V_\w} - \frac{\dot{V}_\w}{V_\w}\frac{L_\w}{V_\w} = \frac{\dot{L}_\w}{V_\w} - \frac{\dot{f}}{f} \rho_\w \ ,\qquad   \dot{\rho}_\c = \frac{\dot{L}_\c}{V_\c} + \frac{\dot{f}}{1-f}\rho_\c \ .
\end{equation}
The density of a dislocation population changes if the length of dislocation segments is altered due to generation or annihilation. Density also changes if there is merely a change of volume occupied by the population. The latter case implies that only the dislocation length is constant, but existing dislocation segments can still move. Thus, the description of cell refinement presented by model \eqref{EstrinOrig1} can be improved. The derivation of \eqref{EstrinOrig1} only consideres the change of dislocation density due to $\dot{L}_\w$ and $\dot{L}_\c$ with a fixed $f$. Thus, the terms $-\dot{f}/f \rho_\w$ and $+\dot{f}/(1-f) \rho_\c$ were ignored. For a physically consistent formulation, the changes due to the evolution of $f$ must be taken into account, too. This becomes especially important if the wall volume rapidly decreases or increases (cf. Section \ref{sec_Reverse}).\par
Considering all the proposed modifications, a refined dislocation density two-phase composite model is obtained:
\begin{subequations}\label{EstrinMod}
\begin{align}
   \doverd{\rho_\w}{s} =& + \frac{6\beta^\ast(1-f)^{2/3}}{bdf}\ \Cut(\rho_\w, \dot{s}) + \frac{\sqrt{3}\beta^\ast(1-f)}{bf}\sqrt{\rho_\w}\notag\\
                        & -k_0\exp\left(- \frac{p\varOmega}{n_\w \kB T}\right) \left( \frac{\dot{s}}{\dot{s}_0}\right)^{\inv /n_\w} \hspace{-5mm} \rho_\w - \frac{1}{f}\doverd{f}{s} \rho_\w \label{Gl_rhow2}\\
   \doverd{\rho_\c}{s} =& - \frac{6\beta^\ast}{bd(1-f)^{1/3}}\ \Cut(\rho_\c, \dot{s}) + \frac{6\alpha^\ast}{\sqrt{3}b}\sqrt{\rho_\w}\notag\\
                        & - k_0\left( \frac{\dot{s}}{\dot{s}_0}\right)^{\inv /n_\c} \hspace{-5mm}\rho_\c  + \frac{1}{1-f}\doverd{f}{s} \rho_\c \quad.\label{Gl_rhoc2}
\end{align}\end{subequations}
Compared to \eqref{EstrinOrig1} it offers the following advantages: 1) The positivity of the dislocation densities is guaranteed. 2) The dependence of the model on the macroscopic state of deformation is unambiguously defined by Eq. \eqref{Defs}. 3)  It is physically consistent in such a way that without generation or annihilation the total dislocation density $\rho_\t$ remains constant.

\section{Parameter identification on cell size measurements}\label{sec_Ident}

The modified two-phase composite model \eqref{EstrinMod}, which has been proposed in the previous section contains a number of parameters (cf. Table \ref{constants}). The fundamental physical constants and processing parameters are known. Further, the material parameters $\alpha, f_0, f_\infty $ defining the evolution of the cell wall fraction $f$ can be determined from TEM measurments \citep{Mckenzie2007}. Hence, seven material constants arranged in the parameter column matrix
\begin{equation}
   \Matr{x}=\left[\alpha^\ast, \beta^\ast, n_\c, n_\w, k_0, \dot{s}_0, K \right]
\end{equation}
are left to be identified. Since they can hardly be measured \emph{directly}, a refined strategy for parameter identification has to be used. Basically, the elements of $\Matr{x}$ can be found by demanding that the model prediction for a \emph{measurable} quantity matches best corresponding experimental data. The modified model \eqref{EstrinMod} predicts the evolution of the densities $\rho_\w, \rho_\c$ of the dislocation populations and thus the total density $\rho_\t$ and cell size $d$ as function of the accumulated plastic strain  $s$. It is well known that dislocation densities are extremely hard to measure. Instead of dislocation densities, the cell size measurements\footnote {Note that the \emph{cell} size is not to be confused with the \emph{grain} size.} can be adopted. Thus, an error can be defined as follows:
\begin{equation}
   r_i=d(\Matr{x}, s_i) - \breve{d}(s_i) \ .
\end{equation}
Here, $d(\Matr{x}, s_i)$ denotes the model prediction, $\breve{d}(s_i)$ stands for the cell size measurements at discrete values $s_i$ of accumulated plastic strain. Now, the optimal set $\Matr{x}_{\text{opt}}$ of parameters within $\mathcal{S}\subset\mathbb{R}^7$ can be found by minimizing a sum of squared errors as target function:
\begin{align} \label{targetfunction}
   \Psi(\Matr{x}_{\text{opt}})=\underset{\Matr{x}\in\mathcal{S}}{\min} \left\{\Psi(\Matr{x})\right\} \ ,\quad \Psi(\Matr{x}) = \sum_{i=1}^N r_i^2(\Matr{x}) \ .
\end{align}
Here, $N$ stands for the number of measurement data points. Introducing the residual $\Matr{r}(\Matr{x})=[r_{1}(\Matr{x}),r_{2}(\Matr{x}),\dotso, r_{N}(\Matr{x})]^\T$, it is possible to write $\Psi(\Matr x)={\Matr r}^\T \Matr r$. As a general requirement, the experimental data (here $\breve{d}$) must provide sufficient information in order to \emph{reliably} identify $\Matr{x}$. The presented model \eqref{EstrinMod} contains parameters which control the response under a change in temperature $T$, hydrostatic pressure $p$ or accumulated plastic strain rate $\dot{s}$, respectively. Therefore, experimental data $\breve{d}(s_i)$ at varying values of $T$, $p$ and $\dot{s}$ is required to determine $n_\c, n_\w, k_0, \dot{s}_0$. The ECAP experiments by \citet{Mckenzie2007} were conducted under different values of hydrostatic pressure by adding a so-called back pressure $p_\text{B}=0\unit{MPa}$ and $p_\text{B}=200\unit{MPa}$. But neither temperature nor strain rate have been varied. Still trying to identify the parameters $n_\c, n_\w, k_0, \dot{s}_0$ would lead to strong correlations among them and most likely poor results. This can be prevented by merging the corresponding parameters:
\begin{align}
   k_\w &\coloneqq k_0\exp\left(-\frac{p\varOmega}{n_\w \kB T}\right) \left( \frac{\dot{s}}{\dot{s}_0}\right)^{\inv /n_\w} \ ,\\
   k_\c &\coloneqq k_0\left( \frac{\dot{s}}{\dot{s}_0}\right)^{\inv /n_\c} \ .
\end{align}
Effectively, we get one single parameter for each dislocation population controling its annihilation rate. This reduces the number of parameters and suppresses strong correlations between them. The simplified evolution equations now read as
\begin{subequations}\label{EstrinModMod}
\begin{align}
   \doverd{\rho_\w}{s} =& + \left(\frac{6\beta^\ast(1-f)^{2/3}}{bKf}\right) \Cut(\rho_\c, \dot{s})\sqrt{\rho_\t} + \left(\frac{\sqrt{3}\beta^\ast(1-f)}{bf}\right)\sqrt{\rho_\w}\notag\\ 
                        & - \left( k_\w  + \frac{1}{f}\doverd{f}{s}\right) \rho_\w \ ,\label{Gl_rhow3}\\
   \doverd{\rho_\c}{s} =& - \left(\frac{6\beta^\ast}{bK(1-f)^{1/3}}\right) \Cut(\rho_\c, \dot{s})\sqrt{\rho_\t} + \left(\frac{6\alpha^\ast}{\sqrt{3}b}\right)\sqrt{\rho_\w} \notag\\ 
                        & - \left(k_\c  - \frac{1}{1-f}\doverd{f}{s}\right) \rho_\c \ .\label{Gl_rhoc3}
\end{align}\end{subequations}
Since $k_\w$ is pressure-dependent, we introduce one such parameter for $p_\text{B}=0\unit{MPa}$ and for $p_\text{B}=200\unit{MPa}$. This way, we are able to consider both data sets from \citet{Mckenzie2007} in the residual but have only one additional parameter. To further reduce the number of parameters, the scaling constant $K$ is also taken from \citet{Mckenzie2007} for $p_\text{B}=0\unit{MPa}$ and $p_\text{B}=200\unit{MPa}$. Finally, a reduced set $\Matr{x}=\left[\alpha^\ast, \beta^\ast, k_{\w_0}, k_{\w_{200}}, k_\c \right]$ has to be identified. It is beneficial to only search within a subset $\mathcal{S}\subset\mathbb{R}^5$ defined by the ranges:
\begin{align}\label{restrict}
   0\leq\alpha^\ast, \beta^\ast \leq1 \quad,\quad k_{\w_0}, k_{\w_{200}}, k_\c \geq 0 \quad .
\end{align}
These restrictions follow immediately from the physical meaning of the constants. For numerical reasons it is favorable when all parameters show a similar order of magnitude. If this is not a priori the case, they can be scaled by characteristic factors $W_j$:
\begin{equation}
    \Matr{x}=\Matr{W}\bar{\Matr{x}}\quad\text{with}\quad \Matr{W}=\diag(W_1,W_2,\dotso) \ .
\end{equation}
The factors can be chosen in such a way that the initial set ${^0}\bar{\Matr{x}}$ only contains values equal to one. Thus, the scaling factors represent the actual initial approximation of the parameters. For the present example the following was set:
\begin{align}\label{startW}
   {^0}\bar{\Matr{x}} = [1,1,1,1,1]^\T \quad\text{and}\quad \Matr{W}=\diag(0.003, 0.006, 3.0, 3.0, 5.0) \ .
\end{align}
Consequently, the sum of squared errors is now a function of the dimensionless\footnote{Note that since the parameters from set $\Matr{x}$ have no dimension, no units appear in the matrix $\Matr{W}$.} and \emph{scaled} parameters: $\Psi=\Psi(\Matr{W}\bar{\Matr{x}})=\Psi(\bar{\Matr{x}})$. To incorporate the restrictions from \eqref{restrict} into \eqref{targetfunction}, an interior-reflective {Newton} algorithm \citep{Coleman1996} with trust region method is applied to solve the minimization problem. A reasonably small target function value was chosen as the abort criterion for the procedure.\par
\begin{figure}[h!]\centering
   \psfrag{A}[m][][1][0]{ Number of ECAP passes }
   \psfrag{B}[m][][1][0]{ $d/\unit{\upmu m}$ }
   \psfrag{C}[l][][1][0]{ simulation for $p_\text{B}=200\MPa$}
   \psfrag{D}[l][][1][0]{ simulation for $p_\text{B}=0$}
   \psfrag{E}[l][][1][0]{ experiment for $p_\text{B}=200\MPa$}
   \psfrag{F}[l][][1][0]{ experiment for $p_\text{B}=0$}
   \includegraphics[scale=0.30]{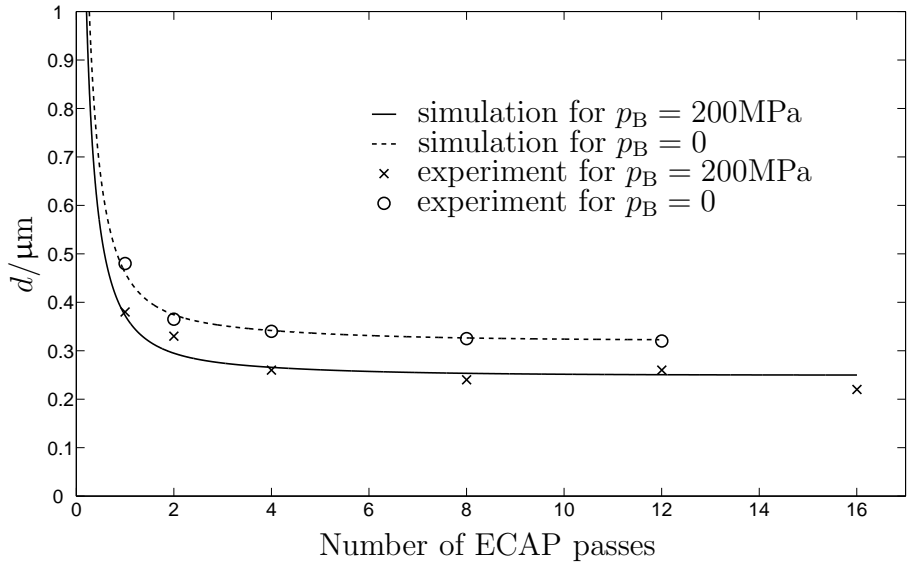}
   \caption[]{Cell size measurements by \citet{Mckenzie2007} for the aluminum alloy AA 6016 and simulated curves using $\Matr{x}_{\text{opt}}$.}\label{IdentPlot}
\end{figure}
Unfortunately, the given problem is not well posed. Depending on the start values (here $\Matr{W}$), different sets of parameters leading to equally suitable minimizations of the target function were found. Eventually, the parameter set $\Matr{x}_{\text{opt}}$ was obtained using the factors from \eqref{startW}:
\begin{equation}\label{finalParams}
   [\alpha^\ast=0.0015, \beta^\ast=0.0017, k_{\w_0}=2.7120, k_{\w_{200}}=2.4526, k_\c=4.9641] \ .
\end{equation}
The fitting result is depicted in Fig \ref{IdentPlot}. Regarding the annihilation parameter we find $k_{\w_0}>k_{\w_{200}}$ which is a reasonable result: in the case of higher hydrostatic pressure, the annihilation rate is smaller due to the pressure dependence of the non-conservative dislocation climb. Note that the annihilation rate $k_\c$ in the cell interior is significantly larger than in the cell wall. This point will be discussed further in Section \ref{sec_FluxWC}.
Even though the number of parameters was reduced to five, strong correlations were still found among them. Nevertheless, the parameters are physically meaningful as will be seen in Section \ref{sec_Simul}.

\section{Extension to non-proportional loading}\label{sec_Reverse}

\subsection{Experimental findings and motivation}\label{ssec_Experimental}

It is experimentally confirmed that cellular dislocation structures may become unstable upon stress reversal \citep{Hasegawa1986}. For pure aluminum and some Al-alloys, \citet{Hasegawa1975} found that cell walls consisting of tangled dislocations are dissolved and become indistinct shortly after a change from tension to compression. At the same time, the total dislocation density significantly decreases and its distribution becomes more uniform. Due to the subsequent monotonic loading, cell patterns emerge again. If this stage of deformation is sufficiently long, the total dislocation density increases and exceeds its previous value while the cell size decreases again. \par
It seems promising to extend the modified two-phase composite model to capture effects due to load path changes. In order to make the model sensitive to such non-proportional deformation, a continuum mechanics technique is required. As a practical tool, the finite strain viscoplasticity model from \citet{Shutov2008a} is used to evaluate the state of strain and stress. In the following, this model is outlined briefly.

\subsection{Finite strain viscoplasticity model}\label{viscoplasticity}

Let us consider the material model of finite strain viscoplasticity, which was proposed by \citet{Shutov2008a}. The kinematics of the model are in essence based on
the double multiplicative split of the deformation gradient
\begin{equation}\label{doublspl}
   \Ten2F=\Fk{\e}\pkt\Fk{\i} \quad,\quad \Fk{\i}=\FKK{\ie}\pkt\FKK{\ii}\ .
\end{equation}
The first decomposition can be motivated by the idea of local elastic unloading. The second decomposition was proposed by \citet{Lion2000} in order
to represent nonlinear kinematic hardening of Armstrong-Frederick type. According to this assumption, the inelastic part $\Fk{\i}$ is decomposed into a conservative part $\FKK{\ie}$ and a dissipative part $\FKK{\ii}$. \citet{Shutov2008a} originally formulated the material model with respect to physically meaningful intermediate configurations. Along with the well-known right {Cauchy-Green} tensor $\Ten2C= {\Ten2F}^\T\cdot\Ten2F$, four tensors of the same type are introduced:
\begin{align}
   \Ci    &= {\Fk{\i}}^\T\cdot\Fk{\i}   \ ,          &\CII     &= {\FKK{\ii}}{^\T}\pkt\FKK{\ii}\ ,\label{subeq4}\\
   \Ce    &= {\Fk{\e}}^\T\cdot\Fk{\e}\ ,          &\vCie &= {\FKK{\ie}}{^\T}\pkt\FKK{\ie}\label{subeq5}\ .
\end{align}
Here, the overset symbols denote the configurations $\hat{\mathcal{K}}$ and $\check{\mathcal{K}}$, where the tensors operate. In addition to the velocity gradient tensor $\Ten2L\coloneqq\Ten2{\dot{F}}\pkt{\Ten2F}^{\inv}$, we define two inelastic counterparts
\begin{equation}\label{GeschGrad}
   \Li=\DotFk{\i}\pkt{\Fk{\i}}^{\inv} \quad\text{and}\quad \vLii=\DotFKK{\ii}\pkt{\FKK{\ii}}^{\en\inv}\ .
\end{equation}
Taking the symmetric parts we obtain the inelastic strain rate and its equivalent for the microstructure:
\begin{equation}
   \Di= \mfrac{1}{2}\left(\Li +\Li^\T \right) \quad,\quad \vDii= \mfrac{1}{2}\left(\vLii + \vLii^\T \right) \ .
\end{equation}
Next, in order to capture the nonlinear isotropic hardening, we split the accumulated inelastic arc length $s$ into a conservative and a dissipative part by $s=s_{\e}+s_{\d}$. The free energy of the viscoplastic body is then decomposed into three independent summands
\begin{equation}\label{ZerlpsiF}
   \psi(\Ce,\vCie,s_{\e})= \psiel(\Ce)+\psi_{\text{kin}}(\vCie)+\psiiso(s_{\e})\ ,
\end{equation}
capturing the macroscopic elastic strain energy $\psiel$ and the energy stored in the microstructure related to kinematic ($\psi_{\text{kin}}$) and isotropic ($\psiiso$) hardening. Now the weighted Cauchy stress tensor acting on the intermediate configuration, the backstress tensor and a scalar stress measure are postulated as
\begin{equation}\label{vorSigma}.
   \hS = 2\tilde{\rho}\frac{\partial\psiel}{\partial\Ce} \quad,\quad 
   \vX = 2\tilde{\rho}\frac{\partial\psikin{}}{\partial\vCie} \quad,\quad R = \tilde{\rho}\frac{\partial\psiiso}{\partial s_{\e}} \ .
\end{equation}
Here $\tilde{\rho}$ denotes the mass density in the reference configuration. Evaluating the {Clausius-Duhem} inequality, we get
\begin{equation}\label{ErstCDU}
   \left(\Ce\pkt\hS-\hX\right)\ppkt\Di+\left(\vCie\pkt\vX\right)\ppkt\vDii-R\dot{s}_{\e} \geq 0 \ ,
\end{equation}
which suggests introducing two ``driving forces'' of the {Mandel} stress tensor type:
\begin{equation}\label{mandel}
   \PI\coloneqq\Ce\pkt\hS-\hX \quad,\quad \vP\coloneqq\vCie\pkt\vX
\end{equation}
with $\hX=\FKK{\ie}\pkt\vX\pkt{\FKK{\ie}}{^{\En\T}}$. The difference between the Mandel stress $\Ce\!\pkt\!\hS$ and the back stress $\hX$ yields the so-called effective stress, which is considered as the driving force of (macroscopic) plastic flow. The inelastic flow on the microstructural level is assumed to be driven by the back stresses. The evolution equations for the state variables are chosen in such a way that the model a priori fullfills the Clausius-Duhem inequality \citep[cf.][]{Shutov2008a}:
\begin{equation}\label{EvuOrig}
   \Di         = \lambda_\i\PI^\D/{\|\PI^\D\|}\quad,\quad
   \vDii       = \lambda_\i\varkappa\vP^\D \quad,\quad
   \dot{s}_{\d}= \lambda_\i\sqrt{\mfrac{2}{3}} \frac{\beta}{\gamma}R\ .
\end{equation}
Here, the notation $\lambda_\i$ stands for the inelastic multiplier, which describes the rate of inelastic deformation. It is determined by the definition of the overstress\footnote{This notation is used to avoid confusion. Originally, \citet{Shutov2008a} denoted the overstresses by $f$.} $\phi$ in combination with {Perzyna}'s rule: 
\begin{align}\label{EvuOrig2}
   \phi\coloneqq\|\PI^\D\|-\sqrt{\mfrac{2}{3}}(\sigma_{\text{F}}+R) \ ,\quad
   \lambda_\i=\frac{1}{\eta}{\left\langle \frac{\phi}{\phi_0}\right\rangle}^m ,\quad \langle x\rangle\coloneqq
   \begin{cases}
      x\ :\ x>0\\
      0\ :\ x\leq0
   \end{cases}\hspace{-2mm} ,
\end{align}
where $\phi_0$ represents the characteristic overstress and $\sigma_{\text{F}}$ stands for the initial uniaxial flow stress. Assuming the isotropic real-valued functions $\psi_{\text{el}}$, $\psi_{\text{kin}}$ and $\psiiso$ to be known and imposing initial conditions, the system of constitutive equations is closed. In the current study, the elastic potential $\psi_{\text{el}}$ corresponds to a compressible Neo-{Hooke} with a bulk modulus $k$ and shear modulus $\mu$, and $\psi_{\text{kin}}$ to the Neo-Hookean ansatz with a modulus $c/2$. By chosing $\psiiso =\gamma\cdot s^{2}_{\e}/2$, the well-known {Voce} rule will be obtained.
The material parameters are $\tilde{\rho} > 0$, $\varkappa \geq 0$, $\beta \geq 0$, $\gamma\in\mathbb{R}$, $\eta \geq 0$, $m \geq 1$, $\sigma_{\text{F}} > 0$, $k > 0$, $\mu > 0$, $c > 0$. \citet{Shutov2008a} transformed the constitutive equations to the reference configuration in order to simplify the numerical treatment. The implementation of the model was discussed by \citet{Shutov2008b, Shutov2010}.

\subsection{Coupling with the refined two-phase composite model}\label{sec_FluxWC}

A common way to measure the intensity of load path changes consists in evaluating the angle between former and current loading direction. Towards that end, first the scalar product and the {Frobenius}-norm of two second-rank tensors $\Ten2U$ and $\Ten2V$ are given as
\begin{equation}\label{defFrob}
    \Ten2U:\Ten2V \coloneqq \Ten2U \ppkt {\Ten2V}^\T= \Sp(\Ten2U\pkt{\Ten2V}^\T) \quad,\quad \|\Ten2U\|\coloneqq \sqrt{\Ten2U:\Ten2U} \ .
\end{equation}
Generalizing the definition of the angle between two first-rank tensors yields a similar invariant for second-rank tensors \citep[cf.][]{Schmitt1985}:
\begin{equation}\label{defAngle}
    \cos\sphericalangle\big(\Ten2U, \Ten2V\big) \coloneqq \frac{\Ten2U : \Ten2V}{\|\Ten2U\| \|\Ten2V\|} = \frac{\Ten2U}{\|\Ten2U\|}\ppkt\frac{\Ten2V^\T}{\|\Ten2V\|}\ .
\end{equation}
\citet{Wilson1994} captured the change of loading direction by evaluating \eqref{defAngle} with the ``old'' plastic strain tensor at the time $t$ and the ``new'' one at $t+\Delta t$. \citet{Yalcinkaya2009} specified \eqref{defAngle} with the inelastic velocity gradient instead:
\begin{equation}\label{ThetaYalc}
    \cos\varTheta = \cos\sphericalangle\big(\Li^{\text{old}}, \Li^{\text{new}}\big) = \frac{\Li^{\text{old}}}{\|\Li^{\text{new}}\| } : \frac{\Li^{\text{new}}}{\|\Li^{\text{old}}\|}\ ,
\end{equation}
whereas \citet{Viatkina2007} uses only the symmetric part of $\Li$, i.e. the inelastic strain rate tensor:
\begin{equation}\label{ThetaViat}
    \cos\varTheta = \cos\sphericalangle\big(\Di^{\text{old}}, \Di^{\text{new}}\big) = \frac{\Di^{\text{old}}}{\|\Di^{\text{new}}\| } \ppkt \frac{\Di^{\text{new}}}{\|\Di^{\text{old}}\|}\ .
\end{equation}
In contrast to \eqref{ThetaYalc}, the inelastic spin does not enter this definition. In the present study, saving old values of $\Di$ can be avoided by exploiting the benefits of the viscoplastic model. As typical for kinematic hardening of {Armstrong-Frederick} type, the back stress tensor lags behind the inelastic strain rate tensor. Thus, $\hX$ automatically saves the previous loading direction and we can postulate the load path change measure as follows:
\begin{equation}\label{Theta}
    \cos\varTheta = \cos\sphericalangle\big(\hX,\Di\big) = \hX\ppkt\Di\ \big/\ \big( \|\hX\| \ \|\Di\| \big) \ .
\end{equation}
Monotonic loading is only present when $\varTheta=0$ whereas proportional non-monotonic loading is characterized by $\varTheta=0$ or $\varTheta=\pi$. In the case of general non-proportional loading, $\varTheta$ can take any value within $[0,\pi]$.
\begin{remark}\label{remPower}
There is a physical interpretation of definition \eqref{Theta}.
The power per unit volume of the back stresses with respect to plastic flow is given as
\begin{equation}\label{power}
    p = \hX\ppkt\Di = \|\hX\| \ \|\Di\| \cos\sphericalangle\big(\hX,\Di\big)
\end{equation}
and arises from the rate of the free energy related to kinematic hardening (cf. \eqref{ErstCDU}):
\begin{equation}\label{dotpsikin}
    \tilde{\rho}\dot{\psi}_{\text{kin}} = - \left(\vCie\pkt\vX\right)\ppkt\vDii + \hX\ppkt\Di \ .
\end{equation}
It follows from \eqref{EvuOrig} that the first summand on the right hand side of \eqref{dotpsikin} is always negative. This loss corresponds to microstructural processes which are dissipative by nature. In constrast to the second summand: if $\hX$ and $\Di$ are ``parallel'', the power according to \eqref{power} is positive and increases the free energy. If they are opposing, no energy is stored, but released/dissipated.
\end{remark}
Definition \eqref{Theta} is advantageous compared to \eqref{ThetaViat} because it does not lead to a single unit impulse immediately after a load path change.
Due to the evolution of the back stress tensor, $\varTheta$ then \emph{steadily} goes back to zero if the subsequent loading is monotonic. Changing the load path is thus modelled with a certain ``inertia''. This is not only numerically favorable but also believed to better capture the intrinsic physical behavior of materials exposed to load path changes. Motivated by the experimental findings explained in Section \ref{ssec_Experimental} an evolution equation for the cell wall volume fraction is formulated.
Experimental observations show that in the case of load path change, the dislocation cell walls become indistinct \citep{Hasegawa1986}. Still, a complete dissolution to a uniform dislocation distribution is rarely observed \citep{Yalcinkaya2009}. Hence, the process can be considered as an increase of wall thickness and its volume fraction. Accordingly, $f$ should rise shortly after the load path change, depending on its intensity. Afterwards, if the deformation stays monotonic for a sufficiently long time, the wall volume fraction $f$ should decrease again. Furthermore, the evolution of $f$ has to fulfill some additional requirements: a) it must be continuous, b) it should saturate under monotonic loading, c) it should include \eqref{fOrig} as a special case and d) $f$ must remain within the range $(0,1)$. An approach which fulfills a) - c) a priori is presented in the current study. It is based on the relation:
\begin{equation}\label{fevol}
   \doverd{f}{t} = \left\{ \delta\ \Resp(\hX,\Di) - \alpha (f - f_\infty) \right\}\dot{s} \quad,\quad f|_{t=0}=f_0 \ .
\end{equation}
The second term on the right hand side stands for the shrinkage of the wall volume fraction $f$ with progressive deformation and would result in Eq. \eqref{fOrig}. The first summand causes a temporary increase of $f$ as a response to load path changes. Here, the parameter $\delta$ controls the amplitude and the function $\Resp(\hX,\Di)\in[0,1]$ determines the intensity and duration of the response. To ensure requirement d), it is sufficient to show that the slope of $f$ is non-positive in the vicinity of $f=1$:
\begin{align}
   \doverd{f}{s}\bigg|_{f=1} & \leq \max\left\{ \delta\ \Resp(\hX,\Di) - \alpha (1 - f_\infty) \right\} = \delta - \alpha (1 - f_\infty) \ ,\notag\\
   \doverd{f}{s}\bigg|_{f=1} & \leq 0 \quad\text{for}\quad  \delta\leq\alpha (1 - f_\infty) \label{restr_delta}\ .
\end{align}
Thus, chosing $\delta\leq\alpha (1 - f_\infty)$ keeps $f$ within the permissible range. The additional parameter can be obtained by measuring the wall thickness prior to and shortly after load path change. The function determining the response to load path changes is assumed in the form: 
\begin{equation}\label{F2def}
   \Resp(\hX,\Di) = \Inten(\varTheta)\ \big\{\|\hX\|\ / \max(\|\hX\|)\big\}^2 \ .
\end{equation}
Here, $\Inten(\varTheta)\in[0,1]$ is used to measure the intensity of load path change as a function of $\varTheta(\hX,\Di)$. The remaining part is explained as follows: Since $\hX$ saturates in case of constant loading direction, its magnitude indicates how long the monotonic phase has lasted. Assuming that the dislocation cell pattern becomes more and more distinct under monotonic deformation, $\|\hX\|/\max(\|\hX\|)\in[0,1]$ measures indirectly how well defined the structure has become. The more distinct it is, the stronger the dissolution and the related increase of the cell wall volume will be\footnote{In case of permanently changing loading directions there will be no response since $\hX$ remains close to zero.}. The back stresses can also be related to the microstructure: Internal stresses of the dislocation cell structure are considered to be partially responsible for the {Bauschinger} effect, which is taken into account in the viscoplastic model by an evolving $\hX$. In case of load reversal these internal stresses support dislocation re-mobilization and thereby pattern disruption. Here these processes are considered as the origin of the energy release described in Remark \ref{remPower}. Since the influence of the back stress tensor is thus twofold, it enters \eqref{F2def} quadratically.
An adequate estimation of the saturation value of the norm of the back stresses is the given by $\max(\|\hX\|) \approx 1/\varkappa$ \citep[cf.][]{Shutov2010b}. $\Inten(\varTheta)$ can be regarded as a ``response function'' for $\|\hX\|\approx \max(\|\hX\|)$ and should be determined according to experimental findings.\par
\begin{figure}[h]\centering
   \psfrag{T}[m][][1][0]{ $\varTheta/\pi$ }
   \psfrag{I}[m][][1][0]{ $\Inten(\varTheta)$}
   \psfrag{Tcrit}[m][][1][0]{\footnotesize $\varTheta_{\text{crit}}$}
   \psfrag{xxxxxxxxxxxxxx}[m][][1][0]{ \footnotesize{piecewise linear} }
   \psfrag{B}[l][][1][0]{ \footnotesize$1-|\cos\varTheta|$ }
   \includegraphics[scale=0.234]{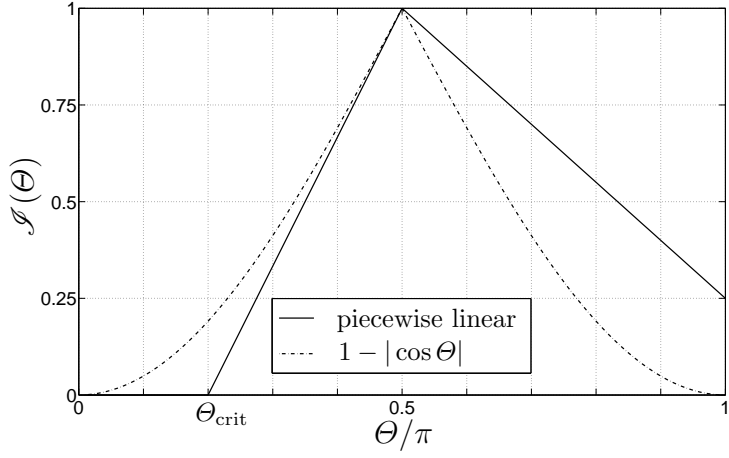}
   \caption[]{Different functions $\Inten(\varTheta)$ measuring the intensity of a load path change.}\label{ResponseFunction}
\end{figure}

 Intuitionally, the strongest response could be expected at the highest value of the load path change measure, i.e. for $\varTheta=\pi$. Such a complete reversal is typical for the {Bauschinger} test. However, experimental findings \citep{Wilson1994,Viatkina2003} suggest the maximum dissolution of cells at so-called cross tests where $\varTheta=\pi/2$, which states a first condition for $\Inten(\varTheta)$. \citet{Viatkina2003} and \citet{Yalcinkaya2009} therefore use the term $1-|\cos\varTheta|$ comparable to the intensity function $\Inten(\varTheta)$ in their modeling.
Two other questions should also be answered to reasonably interpolate the proposed function. Do small load path changes ($\varTheta\rightarrow0$) have an effect? Which intensity $I_{\text{B}}$ is to be expected for the {Bauschinger}-test with $\varTheta=\pi$? Leaving the exact answer open at this stage, we demand from $\Inten(\varTheta)$:
\begin{align}\label{F2conditions}
   \Inten(\varTheta) &= \begin{cases}
                  0                \quad &: \varTheta < \varTheta_{\text{crit}}\\
                  1                \quad &: \varTheta =\pi/2\\
                  \Inten_{\text{B}}\quad &: \varTheta =\pi \end{cases}\\
   \Inten(\pi/2) &> \Inten(\varTheta)\quad\forall\quad\varTheta\neq\pi/2 \ ,
\end{align}
where $\varTheta_{\text{crit}}$ defines a critical load path change which has to be exceeded to trigger any regular cell dissolution. Due to a lack of experimental data, for simplicity we choose $\Inten_{\text{B}}=\frac{1}{4}$, $\varTheta_{\text{crit}}=\pi/5$ and interpolate $\Inten(\varTheta)$ in a piecewise linear way (Fig. \ref{ResponseFunction}).\par
\par

Additionally to the geometrical shift of the boundary (related to a corresponding movement of dislocations) a reverse dislocation flux should to be incorporated. It is known \citep{Amodeo1988} that until the condensation of cell walls to subgrains, the tangled dislocations in the walls can still be activated to become mobile again. Motivated by experimental findings \citep{Christodoulou1986} a load path change is assumed to cause such a re-mobilization and make a fraction of $L_\w$ immigrate to the cell interior. There, fewer obstacles hinder dislocation movement and the chance of mutual annihilation is higher. Effectively, this flux from wall to interior in combination with a higher annihilation rate inside the cells (cf. parameter set \eqref{finalParams}) leads to a more intense decrease in $\rho_\w$. Since, by definition, cell walls are regions of high dislocation density and the magnitude of $\rho_\w$ is at least one order higher than $\rho_\c$, the total dislocation density $\rho_\t$ decreases shortly after a load path change (cf. Eq. \ref{rho_split}). Hence, the modeling of non-proportional loading is in agreement with the experimental oberservations by \citet{Hasegawa1975}. There, it is reported that the total dislocation density decreases shortly after load reversal and the distribution of dislocations becomes more uniform. The model thus gives an explanation for this temporary loss in $\rho_\t$.\par
\begin{figure}[h]\centering
   \psfrag{A}[m][][1][0]{ cell wall }
   \psfrag{B}[m][][1][0]{ interior }
   \psfrag{C}[m][][1][0]{ $d$}
   \psfrag{D}[m][][1][0]{ $v_\w\Delta t$}
   \psfrag{E}[m][][1][0]{ $\dfrac{w}{2}$}
   \includegraphics[scale=0.7]{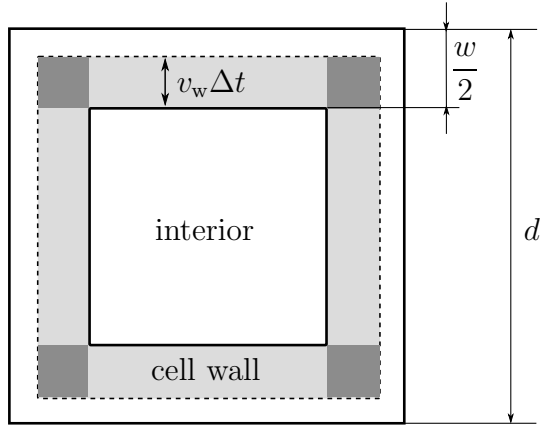}
   \caption[]{Cut of the ``elementary cell'' of the dislocation pattern. The shaded regions stand for the ``collecting area'' of dislocations from the cell walls which are re-mobilized. The dark grey regions can be neglected.}\label{fluxWC}
\end{figure}

The derivation of the flux of dislocations from wall to interior is similar to the reasoning in Section \ref{subsec_Improve}. It is assumed that a fraction $\varGamma^\ast$ of dislocation segments woven in the wall within a distance $v_\w\Delta t$ from the $6$ walls is re-mobilized (Fig. \ref{fluxWC}).
\noindent
That gives a volume of $V_\w^\ast=6 v_\w\Delta t (d-w)^2$, neglecting the parts which are $(\Delta t)^2$ small (dark grey regions in Fig. \ref{fluxWC}). Applying the modified version of {Orowan}'s equation to the re-mobilized fraction of $\rho_\w$, the loss of dislocation length in the cell wall is
\begin{align}
   \Delta L_\w &= -\varGamma^\ast \rho_\w V_\w^\ast \overset{\eqref{modorowan}}{=} -6\varGamma^\ast \Cut(\rho_\w, \dot{s})\Delta t(d-w)^2\dot{s}/b\ ,\notag\\
   \frac{\dot{L}_\w}{V_\w} &\overset{\eqref{deff}}{=}-\frac{6\varGamma^\ast(1-f)^{1/3} \dot{s}}{bdf}\ \Cut(\rho_\w, \dot{s}) \ .
\end{align}
To let this process set in only after load path changes, the quantity $\varGamma^\ast$ has to be a function of $\varTheta$, where $\varTheta$ is defined by \eqref{Theta}. For consistency, we put $\varGamma^\ast=\gamma^\ast \Resp(\hX,\Di)$ with the constant $0\leq\gamma^\ast\leq1$. Now the model is complete and the final set of evolution equations for the three state variables is given by
\begin{subequations}\label{EstrinFin}
\begin{align}
   \doverd{\rho_\w}{s} = & +B(\rho_\w,\rho_\c,\hX,\Di)\ \frac{6(1-f)^{\frac{2}{3}}}{bdf}\  + \frac{\sqrt{3}\beta^\ast(1-f)}{bf}\sqrt{\rho_\w}\notag\\
                         & -k_0\e^{\left(- \frac{p\varOmega}{n_\w \kB T}\right)} \left( \frac{\dot{s}}{\dot{s}_0}\right)^{\inv /n_\w} \hspace{-5mm} \rho_\w - \frac{1}{f}\doverd{f}{s} \rho_\w \ ,\\
   \doverd{\rho_\c}{s} = & -B(\rho_\w,\rho_\c,\hX,\Di)\ \frac{6\beta^\ast}{bd(1-f)^{\frac{1}{3}}} + \frac{6\alpha^\ast}{\sqrt{3}b}\sqrt{\rho_\w}\notag\\
                         & -k_0\left( \frac{\dot{s}}{\dot{s}_0}\right)^{\inv /n_\c} \hspace{-5mm}\rho_\c  + \frac{1}{1-f}\doverd{f}{s} \rho_\c \ ,\\
   \doverd{f}{s}       = & +\delta\ \Resp(\hX,\Di) - \alpha (f - f_\infty)\ , \\
                     B = &  \Big\{\beta^\ast \Cut(\rho_\c, \dot{s}) - \gamma^\ast \Cut(\rho_\w, \dot{s}) \Resp(\hX,\Di)\Big\}\ ,\notag\\
                         &\rho_\w|_{s=0}={\rho}_{\w0} \ ,\ \rho_\c|_{s=0}={\rho}_{\c0} \ ,\ f|_{s=0}=f_0\ . 
\end{align}\end{subequations}
In the following this model will be referred to as ``load path sensitive two-population dislocation cell model''. Note that choosing $\Cut=1, \delta=\gamma^\ast=0$ and provided $\dot{f}\rightarrow0$, the original model is included as special case of system \eqref{EstrinFin}.

\section{Simulation results}\label{sec_Simul}

\subsection{Simulation of common Equal Channel Angular Pressing routes}

The goal of the subsequent simulation is to estimate the influence of load path changes on the cell size using the example of Equal Channel Angular Pressing (ECAP). The most simple operation mode of the ECAP process is given by route $A$ \citep{Iwahashi1998}. In this case, the billet is pressed through the anglular channel over and over in the same fashion. Neglecting any curvature of the channel and heterogenity of the resulting deformation, the material flow in the active zone corresponds to a simple shear. Hence, for the inelastic deformation gradient and the inelastic strain rate we get:
\begin{align}
   \Big[F_{\i_{ab}}\Big] =
   \begin{bmatrix*}[c]
    1 & 0& \gamma(t) \\
    0 & 1   & 0 \\
    0 & 0   & 1 \\
\end{bmatrix*}\quad\overset{\eqref{DefD}}{\Rightarrow}\quad
    \Big[\hat{D}_{\i_{ab}}\Big] =\frac{\dot{\gamma}}{2}
   \begin{bmatrix*}[c]
    0 & 0 & 1 \\
    0 & 0 & 0 \\
    1 & 0 & 0 \\
\end{bmatrix*}
\end{align}
with the monotonic function $\gamma(t)$. For the sake of simplicity, the measure \eqref{ThetaViat} is used in this section to estimate the load path change. According to this definition, the angle between $\Di^{\text{new}}$ and $\Di^{\text{old}}$ remains $0$. Clearly, route $A$ represents successive monotonic loading.\par
Let us now recall the load path which corresponds to route $C$ \citep{Iwahashi1998}. After each pass the billet is rotated through $180^\circ$ about its longitudinal axis before it is inserted into the input channel again. Accordingly, there is a simple shear in the $x-z$ plane with the shear angle increasing at each odd pass and decreasing at each even pass. For the inelastic strain rate tensor in the ``old'' $(2N-1)$\textsuperscript{th} and ``new'' $2N$\textsuperscript{th} ECAP pass we obtain
\begin{align}
   \Big[\hat{D}_{\i_{ab}}^{\text{old}}\Big] =+\frac{\dot{\gamma}}{2}
   \begin{bmatrix*}[c]
    0 & 0 & 1 \\
    0 & 0 & 0 \\
    1 & 0 & 0 \\
\end{bmatrix*}\qquad,\qquad
   \Big[\hat{D}_{\i_{ab}}^{\text{new}}\Big] =-\frac{\dot{\gamma}}{2}
   \begin{bmatrix*}[c]
    0 & 0 & 1 \\
    0 & 0 & 0 \\
    1 & 0 & 0 \\
\end{bmatrix*}\ .
\end{align}
Effectively, this corresponds to a complete load reversal, in other words a {Bauschinger}-test with $\varTheta=\pi$:
\begin{equation}
    \cos\varTheta = \Di^{\text{old}}\ppkt\Di^{\text{new}}\ \big/\ \big( \|\Di^{\text{old}}\| \ \|\Di^{\text{new}}\| \big)\ = -1 \ .
\end{equation}
The loading program which corresponds to route $B_c$ is somewhat more complex \citep{Iwahashi1998}. After each pass the billet is rotated through $90^\circ$ around its longitudinal axis. Consequently, there is a simple shear in the $x-z$ plane in the first pass, whereas in the second pass it is the $x-y$ plane. In the third pass the deformation of the first pass is reversed and in the fourth pass the same is done for the shear from the second pass. Then, the procedure starts over. The coefficients of the inelastic strain rate tensors for pass 1 to 4 are summarized in the following table:
\begin{table}[h]
\begin{tabular}{| >{$}c<{$} | >{$}c<{$} | >{$}c<{$} | >{$}c<{$} | >{$}c<{$} | }
\firsthline
N  & 1 & 2 & 3 & 4  \\ \hline\hline
\Big[\hat{D}_{\i_{ab}}\Big]   &
\frac{\dot{\gamma}}{2}
   \begin{bmatrix*}[c]
    0 & 0 & 1 \\
    0 & 0 & 0 \\
    1 & 0 & 0 \\
   \end{bmatrix*}             &
   +\frac{\dot{\gamma}}{2}
   \begin{bmatrix*}[c]
    0 & 1 & 0 \\
    1 & 0 & 0 \\
    0 & 0 & 0 \\
\end{bmatrix*}                &
   -\frac{\dot{\gamma}}{2}
   \begin{bmatrix*}[c]
    0 & 0 & 1 \\
    0 & 0 & 0 \\
    1 & 0 & 0 \\
\end{bmatrix*}                &
   -\frac{\dot{\gamma}}{2}
   \begin{bmatrix*}[c]
    0 & 1 & 0 \\
    1 & 0 & 0 \\
    0 & 0 & 0 \\
\end{bmatrix*}\\ \hline
\end{tabular}
\end{table}\par
The angles between the corresponding $\Di^{\text{new}}$ and $\Di^{\text{old}}$ follow as:
$$
   \cos\overset{12}{\varTheta} = \cos\overset{23}{\varTheta} = \cos\overset{34}{\varTheta} = \cos\overset{41}{\varTheta} = 0 \ .
$$
Thus, we obtain a sequence of cross tests where $\varTheta=\pi/2$. The loading program which corresponds to route $E$ is similar \citep{Barber2004}. There is a simple shear both in the $x-z$ and $x-y$ plane. In contrast to route $B_c$ the deformation of each pass is immediately reversed in the following pass. After four passes, the procedure starts over. The coefficients of the inelastic strain rate tensors for pass 1 to 4 are summarized in the following table:\par
\begin{table}[h]
\begin{tabular}{| >{$}c<{$} | >{$}c<{$} | >{$}c<{$} | >{$}c<{$} | >{$}c<{$} | }
\firsthline
N  & 1 & 2 & 3 & 4  \\ \hline\hline
\Big[\hat{D}_{\i_{ab}}\Big]   & 
   +\frac{\dot{\gamma}}{2}
   \begin{bmatrix*}[c]
    0 & 0 & 1 \\
    0 & 0 & 0 \\
    1 & 0 & 0 \\
   \end{bmatrix*}             &
   -\frac{\dot{\gamma}}{2}
   \begin{bmatrix*}[c]
    0 & 0 & 1 \\
    0 & 0 & 0 \\
    1 & 0 & 0 \\
\end{bmatrix*}                &
   +\frac{\dot{\gamma}}{2}
   \begin{bmatrix*}[c]
    0 & 1 & 0 \\
    1 & 0 & 0 \\
    0 & 0 & 0 \\
\end{bmatrix*}                &
   -\frac{\dot{\gamma}}{2}
   \begin{bmatrix*}[c]
    0 & 1 & 0 \\
    1 & 0 & 0 \\
    0 & 0 & 0 \\
\end{bmatrix*}\\ \hline
\end{tabular}
\end{table}
The angles between the corresponding $\Di^{\text{new}}$ and $\Di^{\text{old}}$ follow as:
$$
   \cos\overset{12}{\varTheta} = \cos\overset{34}{\varTheta} = -1 ,\enspace \cos\overset{23}{\varTheta} = \cos\overset{41}{\varTheta} = 0 \ .
$$
Hence, we obtain a sequence of alternating {Bauschinger} ($\varTheta=\pi$) and cross tests ($\varTheta=\pi/2$).\par
\begin{figure}[h]\centering
   \psfrag{FXY}[m][][1][0]{ $F_{xy}$ }
   \psfrag{FXZ}[m][][1][0]{ $F_{xz}$ }
   \psfrag{N}[m][][1][0]{$N$}
   \includegraphics[scale=0.28]{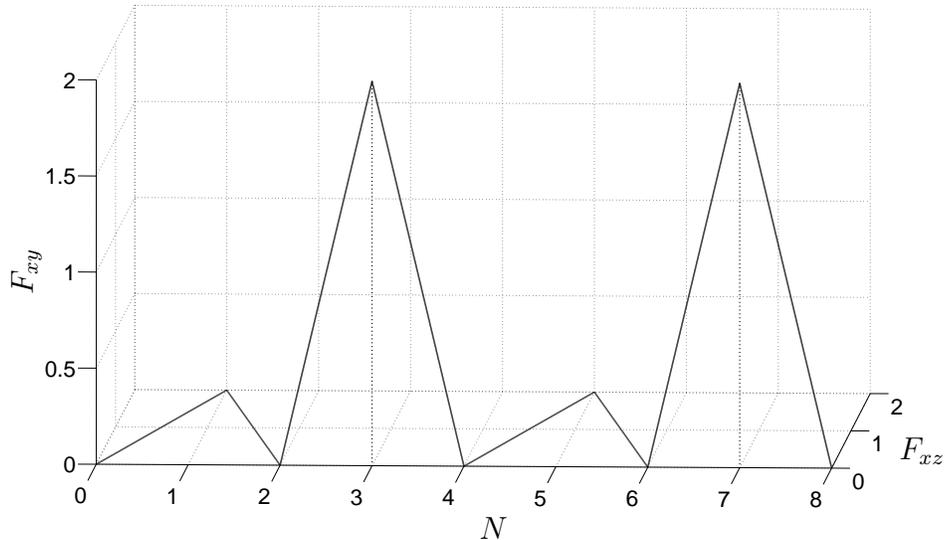}
   \caption[]{Loading program for ECAP route $E$ with $N$ representing the number of ECAP passes.}\label{DefGrECAP_E}
\end{figure}

 
Even though the kinematic considerations of different ECAP routes are idealized, they are useful for developing awareness of basic load path changes of route $A$, $C$, $B_c$ and $E$. Neglecting small elastic deformations such that $\Fk{\e}\approx\Ten2I$ and $\Fk{\i}\approx\Ten2F$, all cases are included in the loading program controlled by the deformation gradient
\begin{align}\label{DefGradECAP}
   \Big[F_{\i_{ab}}\Big] \approx \Big[F_{ab}\Big] =
   \begin{bmatrix*}[c]
    1 & F_{xy}(t) & F_{xz}(t) \\
    0 & 1   & 0 \\
    0 & 0   & 1 \\
\end{bmatrix*}\ .
\end{align}
Here, rigid body rotations of the billet are ignored. Thus, any complex ECAP process is reduced to a sequence of shears applied on different shear planes. Assuming additionally a right-angled channel, a final shear strain of $\max (F_{xy}) = \max (F_{xz})=2$ is reached in each step. For simplicity, the functions $F_{xz}(t)$ and $F_{xy}(t)$ are piecewise linear. Now, the load path changes can be visualized in the $(F_{xz}, F_{xy})$-space. For route $A$ and $C$, $F_{xy}=0$ and the loading history can be visualized in one plane.
\begin{figure}[h]\centering
   \psfrag{FXY}[m][][1][0]{ $F_{xy}$ }
   \psfrag{FXZ}[m][][1][0]{ $F_{xz}$ }
   \psfrag{N}[m][][1][0]{$N$}
   \includegraphics[scale=0.28]{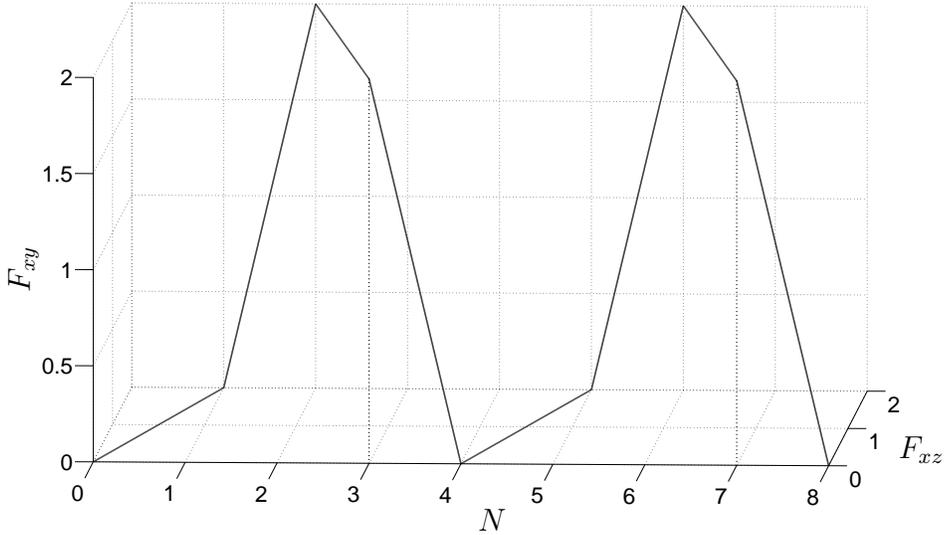} 
   \caption[]{Loading program for ECAP route $B_c$ with $N$ representing the number of ECAP passes.}\label{DefGrECAP_BC}
\end{figure}
The load path changes caused by route $B_c$ and $E$ are illustrated in Fig. \ref{DefGrECAP_BC} and \ref{DefGrECAP_E}, respectively. Here, load path changes are clearly visible as kinks in $(F_{xz}, F_{xy})$-space.
\par\indent
All four idealized ECAP routes were used to evaluate the load path sensitive two-population dislocation cell model described by \eqref{EstrinFin} coupled with the macroscopic viscoplastic model from Section \ref{viscoplasticity}. Since the deformation was assumed to be homogeneous, it suffices to solve the problem at one single material point. For a given local deformation history $\Ten2F(t)$ according to \eqref{DefGradECAP}, the macro and micro material response in the time interval $t \in [0,T]$ was obtained evaluating the viscoplasticity and two-population-dislocation model simultaneously. The material parameters which were used are summarized in Tables \ref{tab_macro} and \ref{tab_micro}.\par
\begin{table}[h]
\caption{Macroscopical material parameters used for numerical simulations}\label{tab_macro}
\begin{tabular}{| >{$}c<{$}| >{$}c<{$} | >{$}c<{$}| >{$}c<{$} | >{$}c<{$}| >{$}c<{$} | >{$}c<{$}| >{$}c<{$} | >{$}c<{$} |}
\firsthline
k/\MPa    & \mu/\MPa  & c/\MPa & \gamma/\MPa &\sigma_{\text{F}}/\MPa & m   & \eta/\unit{s} & \varkappa/\MPa^\inv & \beta \\ \hline\hline
73500     & 28200     & 200   & 460         & 270                   & 3.6 & 2\cdot10^6  &  0.005              & 5     \\ \hline
\end{tabular} \\
\end{table}
The physical constants of the micromodel were taken from Table \ref{constants}. The initial and asymptotic value $f_0,f_\infty$ of the cell wall volume fraction and the scaling constant $K$ were taken from the measurements in \citet{Mckenzie2007}. Since the parameter $\alpha$ is not given there, it was estimated in such a way that $f$ reaches approximately $f_\infty$ after $8$ ECAP passes for Route $A$. Due to a lack of experimental data, the parameter $\delta$ could only be estimated exploiting the restriction imposed by \eqref{restr_delta}. For the same reason the re-mobilization parameter $\gamma^\ast$ was adjusted in such a way as to obtain reasonable results. All remaining parameters were provided by the identified set $\Matr{x}_{\text{opt}}$ from Section \ref{sec_Ident}. Since the simulation was performed assuming constant temperature $T$, hydrostatic pressure $p$ and plastic strain rate $\dot{s}$, the merged annihilation parameters could be used again. No additional back pressure was applied such that $k_{\w}=k_{\w_0}$.
\begin{table}[h]
\caption{Microscopical material parameters used for numerical simulations}\label{tab_micro}
\begin{tabular}{| >{$}c<{$}| >{$}c<{$} | >{$}c<{$}| >{$}c<{$} | >{$}c<{$}| >{$}c<{$} | >{$}c<{$} |}
\firsthline
\alpha^\ast & \beta^\ast & \gamma^\ast & \alpha & \delta  & k_\w   & k_\c    \\ \hline\hline
0.0015      & 0.0017     & 0.0500      & 0.3500 &  0.0700 & 2.7120 & 4.9641  \\ \hline
\end{tabular} \\
\begin{tabular}{| >{$}c<{$} | >{$}c<{$} | >{$}c<{$} | >{$}c<{$} |}
\firsthline
 K      & v_{\text{max}}/\unitfrac{m}{s} & f_0      & f_\infty \\ \hline\hline
 6.2000 & 3100.0                   & 0.0700   & 0.0500   \\ \hline
\end{tabular} 
\end{table}\par
\begin{figure}[h!]\centering
   \psfrag{rho}[m][][1][0]{ $\rho/\unit{m^{\minus2}}$ }
   \psfrag{s}[m][][1][0]{ $s$ }
   \psfrag{rhow}[m][][1][0]{ $\rho_\w$ }
   \psfrag{rhot}[m][][1][0]{ $\rho_\t$ }
   \psfrag{rhoc}[m][][1][0]{ $\rho_\c$ }
   \includegraphics[scale=0.25]{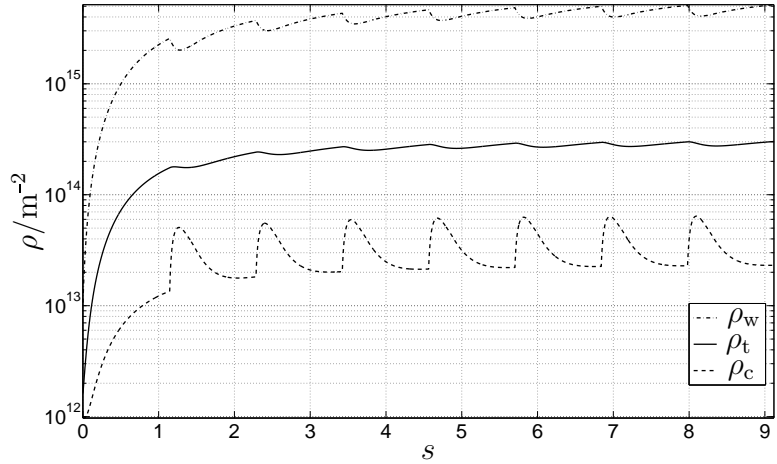}
   \caption[xxx]{Semi-logarithmic plot of the evolution of dislocation populations simulating ECAP Route $C$.}\label{rwrcrtC}
\end{figure}
\begin{figure}[h]
\begin{minipage}{0.5\textwidth}\centering
   \psfrag{f}[m][][1][0]{ $f$ }
   \psfrag{s}[m][][1][0]{ $s$ }
   \includegraphics[width = 0.7\textwidth]{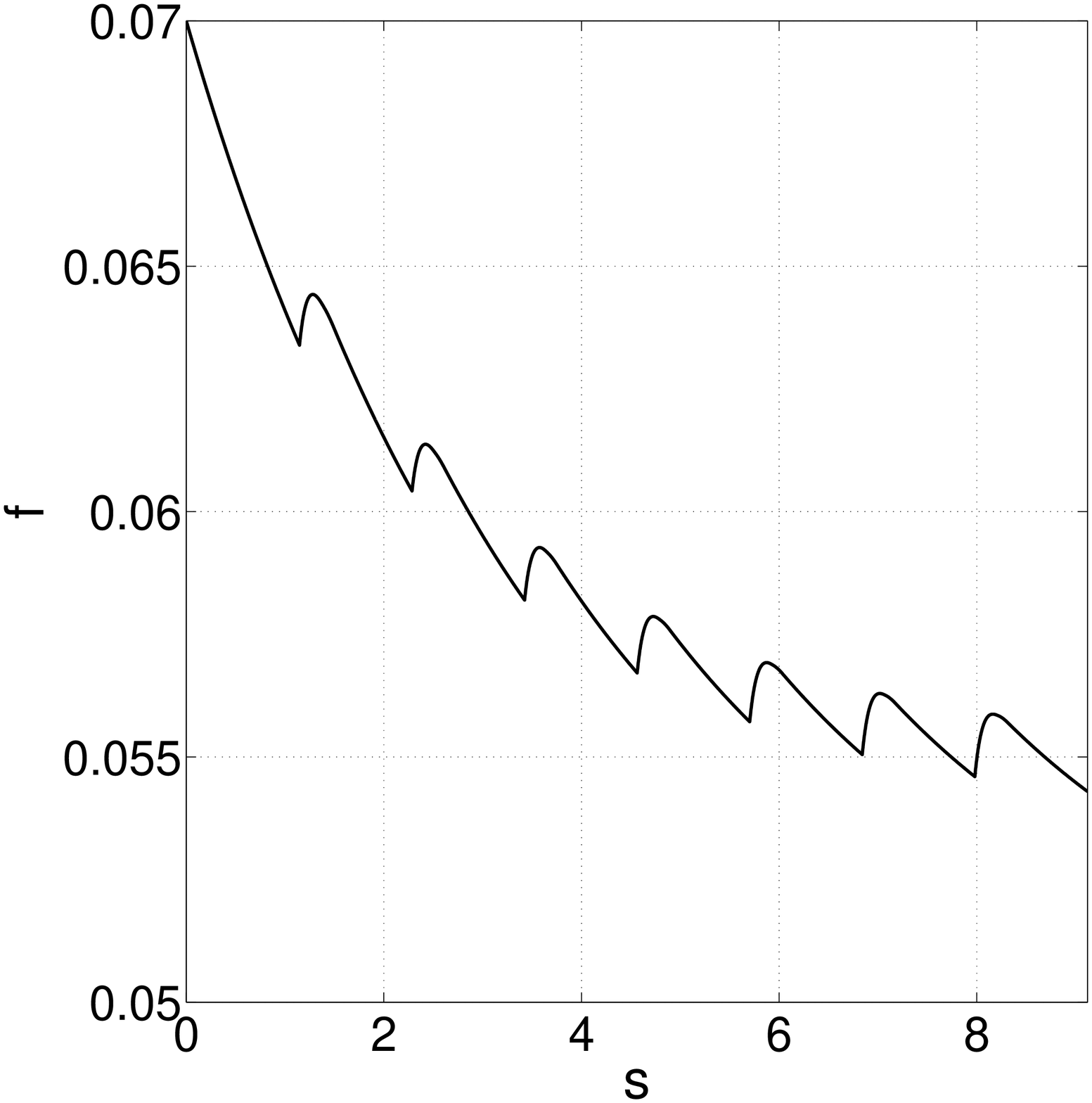}
\end{minipage}
\hspace{-5mm}
\begin{minipage}{0.5\textwidth}\centering
   \psfrag{d}[m][][1][0]{ $d/\unit{\upmu m}$ }
   \psfrag{s}[m][][1][0]{ $s$ }
   \includegraphics[width = 0.7\textwidth]{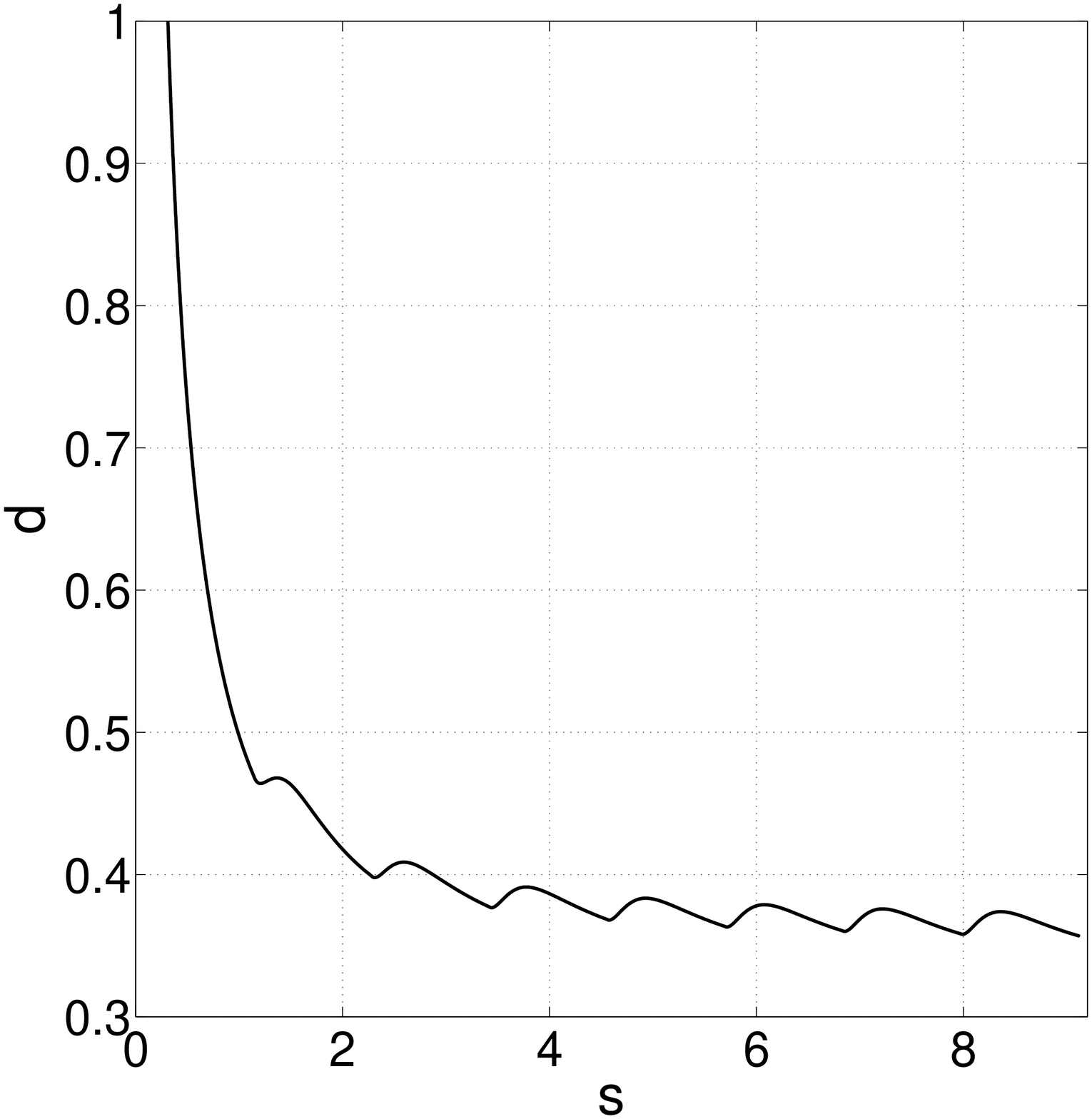}
\end{minipage}
\caption[xxx]{Evolution of the cell wall volume fraction $f$ (left) and size $d$ (right) for ECAP Route $C$ corresponding to Fig. \ref{rwrcrtC}.}\label{fC}
\end{figure}
\begin{figure}[h]\centering
   \psfrag{rho}[m][][1][0]{ $\rho/\unit{m^{\minus2}}$ }
   \psfrag{s}[m][][1][0]{ $s$ }
   \psfrag{rhow}[m][][1][0]{ $\rho_\w$ }
   \psfrag{rhot}[m][][1][0]{ $\rho_\t$ }
   \psfrag{rhoc}[m][][1][0]{ $\rho_\c$ }
   \includegraphics[scale=0.25]{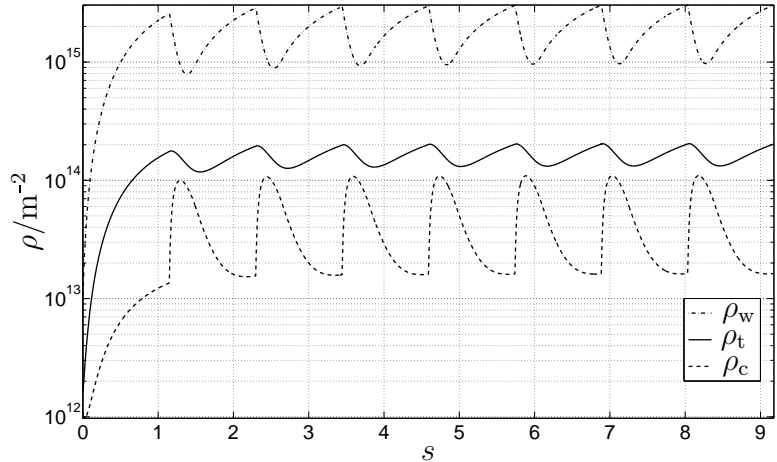}
   \caption[]{Semi-logarithmic plot of the evolution of dislocation populations simulating ECAP Route $B_c$.}\label{rwrcrtB}
\end{figure}
\begin{figure}[h]
\begin{minipage}{0.5\textwidth}\centering
   \psfrag{f}[m][][1][0]{ $f$ }
   \psfrag{s}[m][][1][0]{ $s$ }
   \includegraphics[width = 0.7\textwidth]{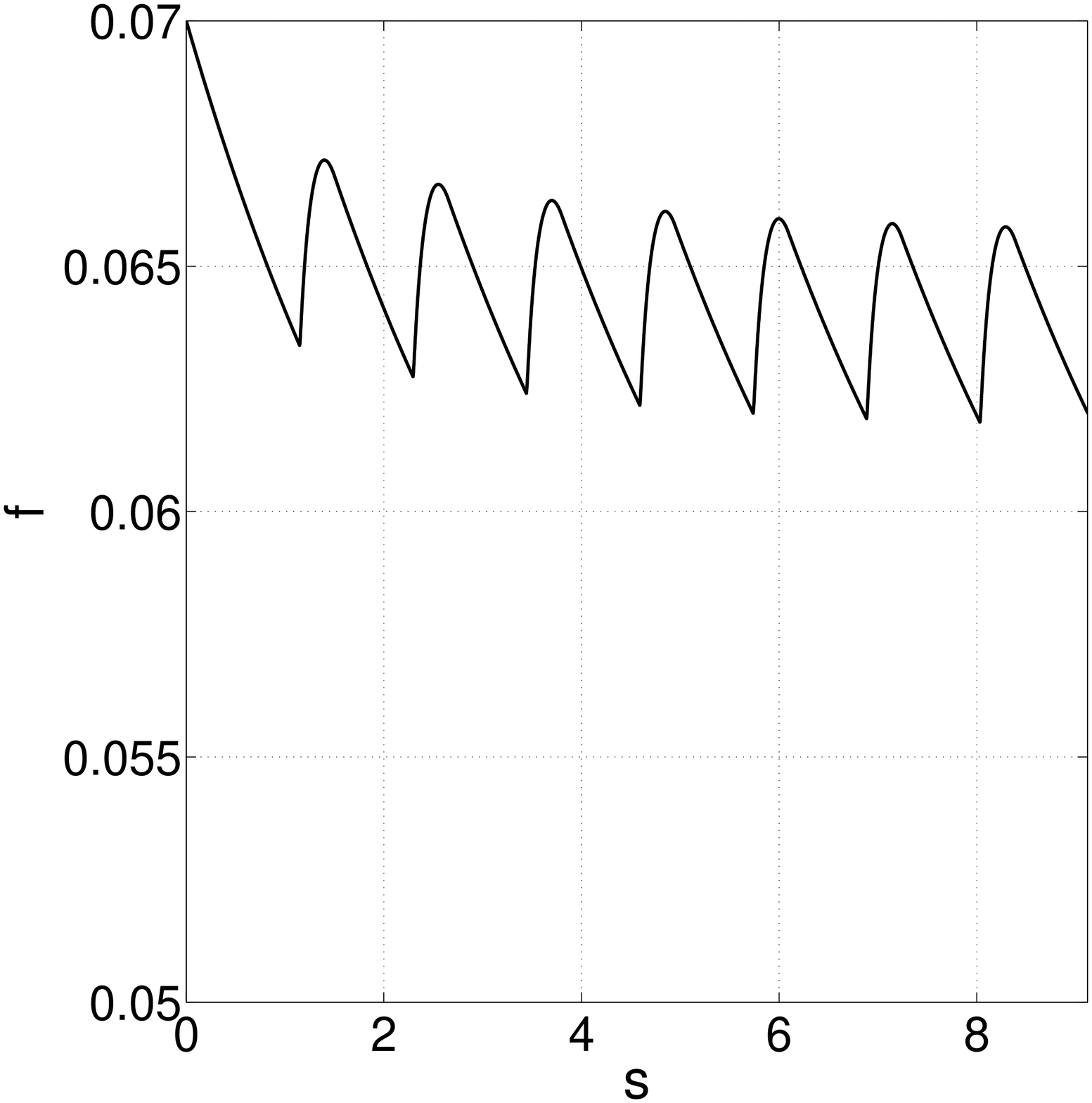}
\end{minipage}
\hspace{-5mm}
\begin{minipage}{0.5\textwidth}\centering
   \psfrag{d}[m][][1][0]{ $d/\unit{\upmu m}$ }
   \psfrag{s}[m][][1][0]{ $s$ }
   \includegraphics[width = 0.7\textwidth]{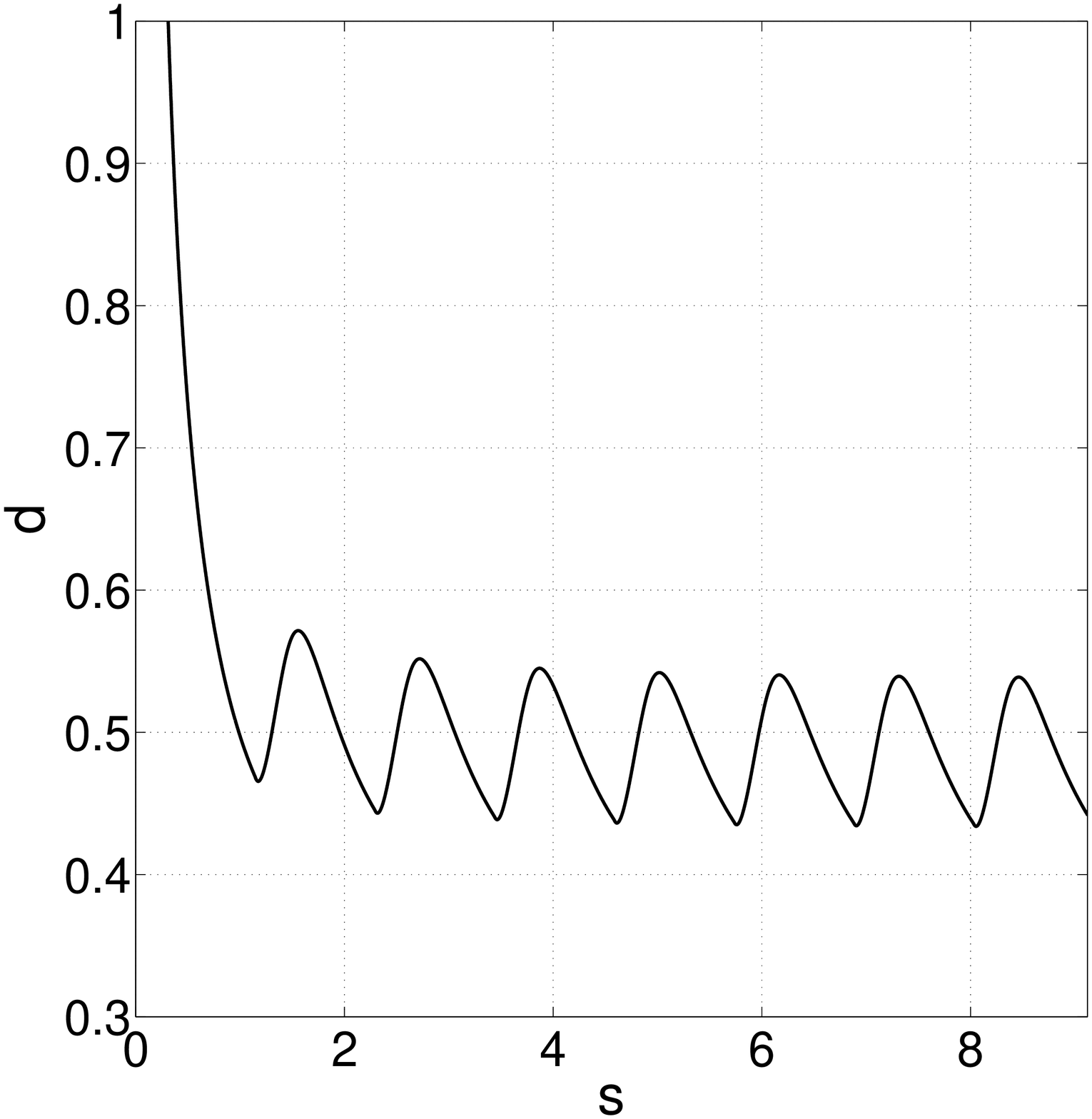}
\end{minipage}
\caption[]{Evolution of the cell wall volume fraction $f$ (left) and size $d$ (right) for ECAP Route $B_c$ corresponding to Fig. \ref{rwrcrtB}.}\label{fB}
\end{figure}
\begin{figure}[h]\centering
   \psfrag{rho}[m][][1][0]{ $\rho/\unit{m^{\minus2}}$ }
   \psfrag{s}[m][][1][0]{ $s$ }
   \psfrag{rhow}[m][][1][0]{ $\rho_\w$ }
   \psfrag{rhot}[m][][1][0]{ $\rho_\t$ }
   \psfrag{rhoc}[m][][1][0]{ $\rho_\c$ }
   \includegraphics[scale=0.25]{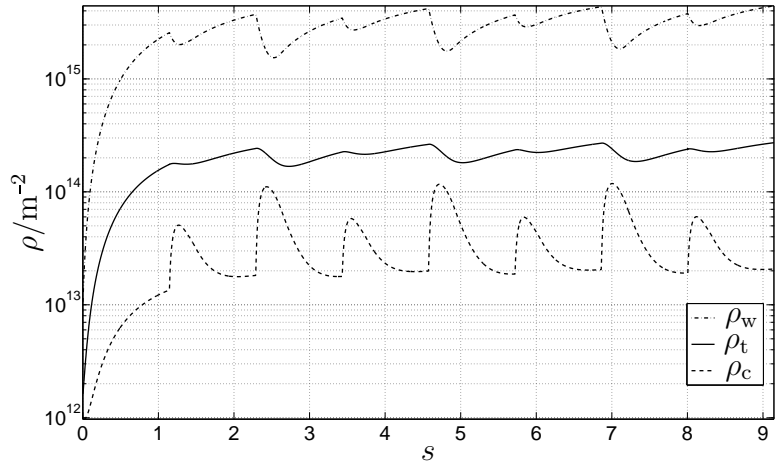}
   \caption[]{Semi-logarithmic plot of the evolution of dislocation populations simulating ECAP Route $E$.}\label{rwrcrtE}
\end{figure}
\begin{figure}[h]
\begin{minipage}{0.5\textwidth}\centering
   \psfrag{f}[m][][1][0]{ $f$ }
   \psfrag{s}[m][][1][0]{ $s$ }
   \includegraphics[width = 0.7\textwidth]{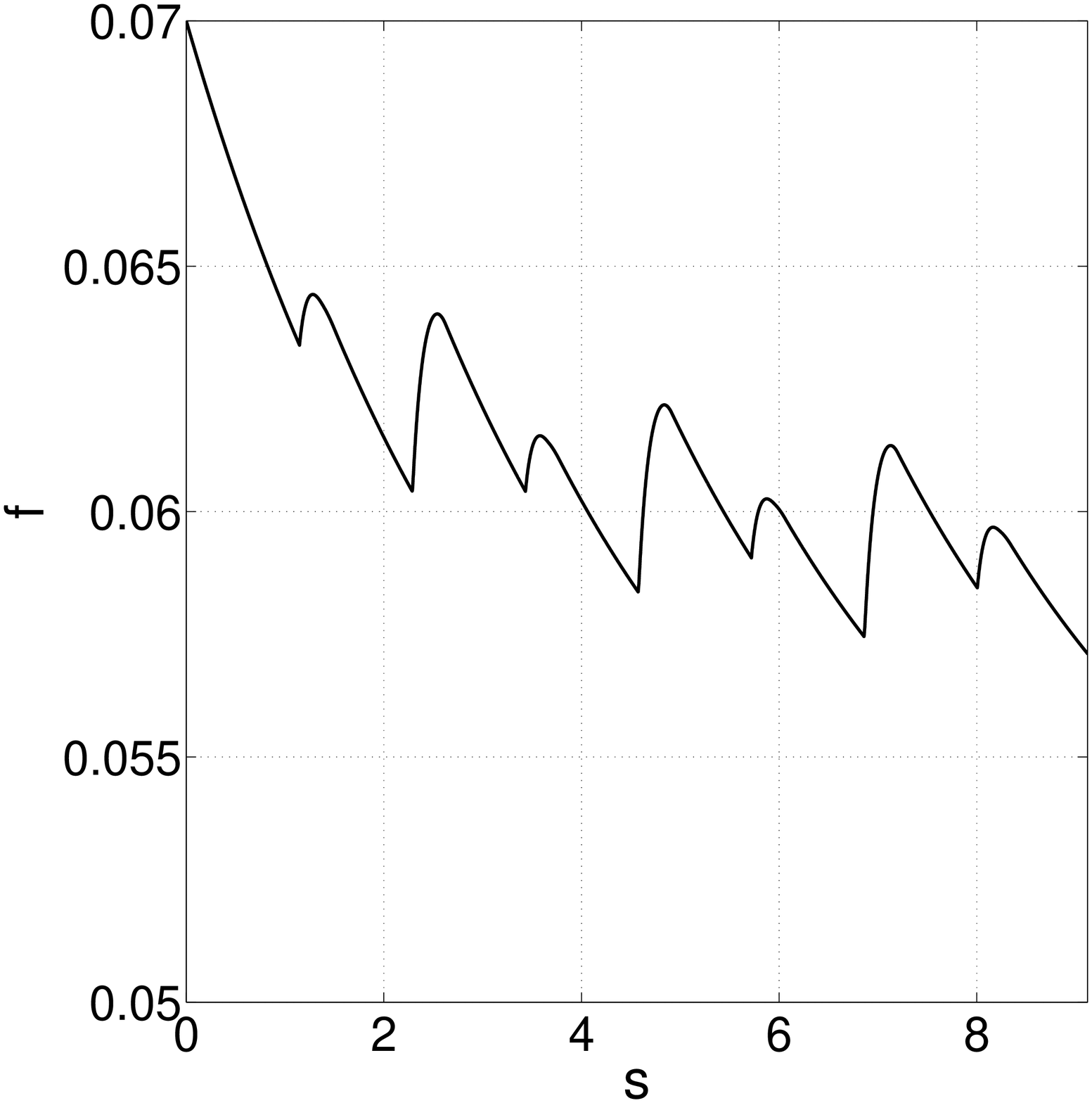}
\end{minipage}
\hspace{-5mm}
\begin{minipage}{0.5\textwidth}\centering
   \psfrag{d}[m][][1][0]{ $d/\unit{\upmu m}$ }
   \psfrag{s}[m][][1][0]{ $s$ }
   \includegraphics[width = 0.7\textwidth]{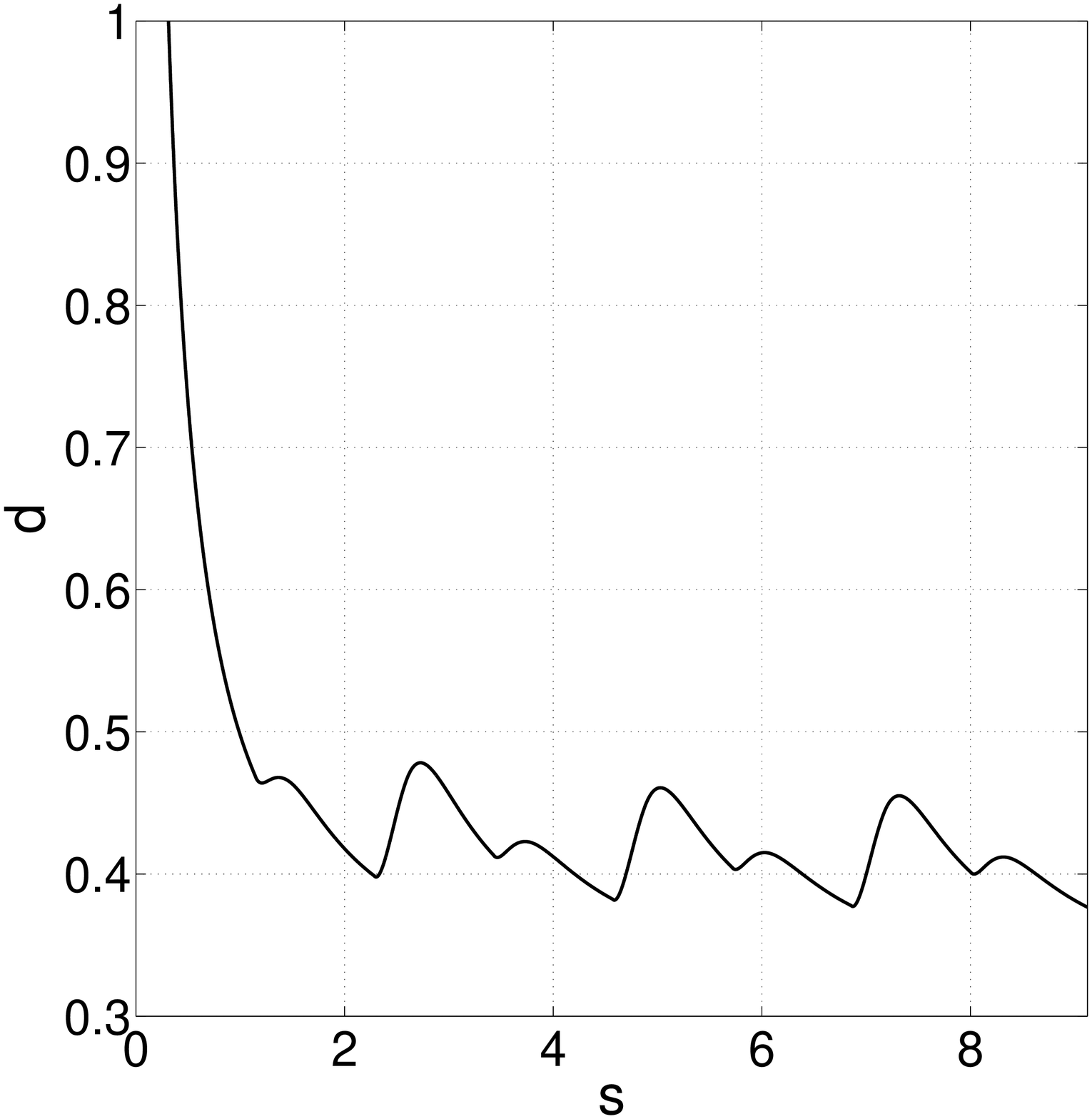}
\end{minipage}
\caption[]{Evolution of the cell wall volume fraction $f$ (left) and size $d$ (right) for ECAP Route $E$ corresponding to Fig. \ref{rwrcrtE}.}\label{fE}
\end{figure}

The initial values $\rho_\w|_{t=0}=10^{13}/\unit{m}^2$ and $\rho_\c|_{t=0}=10^{12}/\unit{m}^2$ were estimated in the following way. The resulting total dislocation density corresponds to a typical value for well-annealed undeformed crystals and the difference between $\rho_\w$ and $\rho_\c$ was supposed to be one order of magnitude.
\begin{figure}[h]\centering
   \psfrag{rho}[m][][1][0]{ $\rho/\unit{m^{\minus2}}$ }
   \psfrag{s}[m][][1][0]{ $s$ }
   \psfrag{rhow}[m][][1][0]{ $\rho_\w$ }
   \psfrag{rhot}[m][][1][0]{ $\rho_\t$ }
   \psfrag{rhoc}[m][][1][0]{ $\rho_\c$ }
   \includegraphics[scale=0.25]{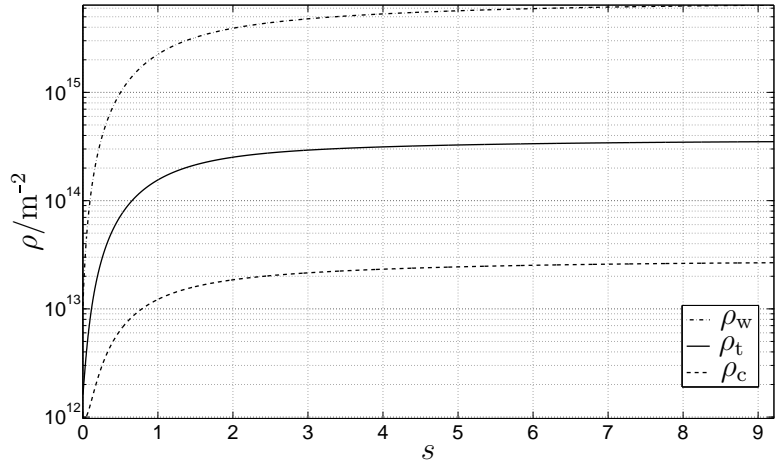}
   \caption[]{Semi-logarithmic plot of the evolution of dislocation populations simulating ECAP Route $A$.}\label{rwrcrtA}
\end{figure}
The simulation results are presented in Fig. \ref{rwrcrtA} - \ref{fE}. The obtained evolution of the dislocation densities for route $A$ serves as reference since it corresponds to monotonic loading. In that case the cell wall volume fraction $f$ decreases monotonically. Both dislocation populations and with them the cell size $d=K/\sqrt{\rho_\t}$ tend to saturate (cf. Fig. \ref{fA}).
\begin{figure}[h]
\begin{minipage}{0.5\textwidth}\centering
   \psfrag{f}[m][][1][0]{ $f$ }
   \psfrag{s}[m][][1][0]{ $s$ }
   \includegraphics[width = 0.7\textwidth]{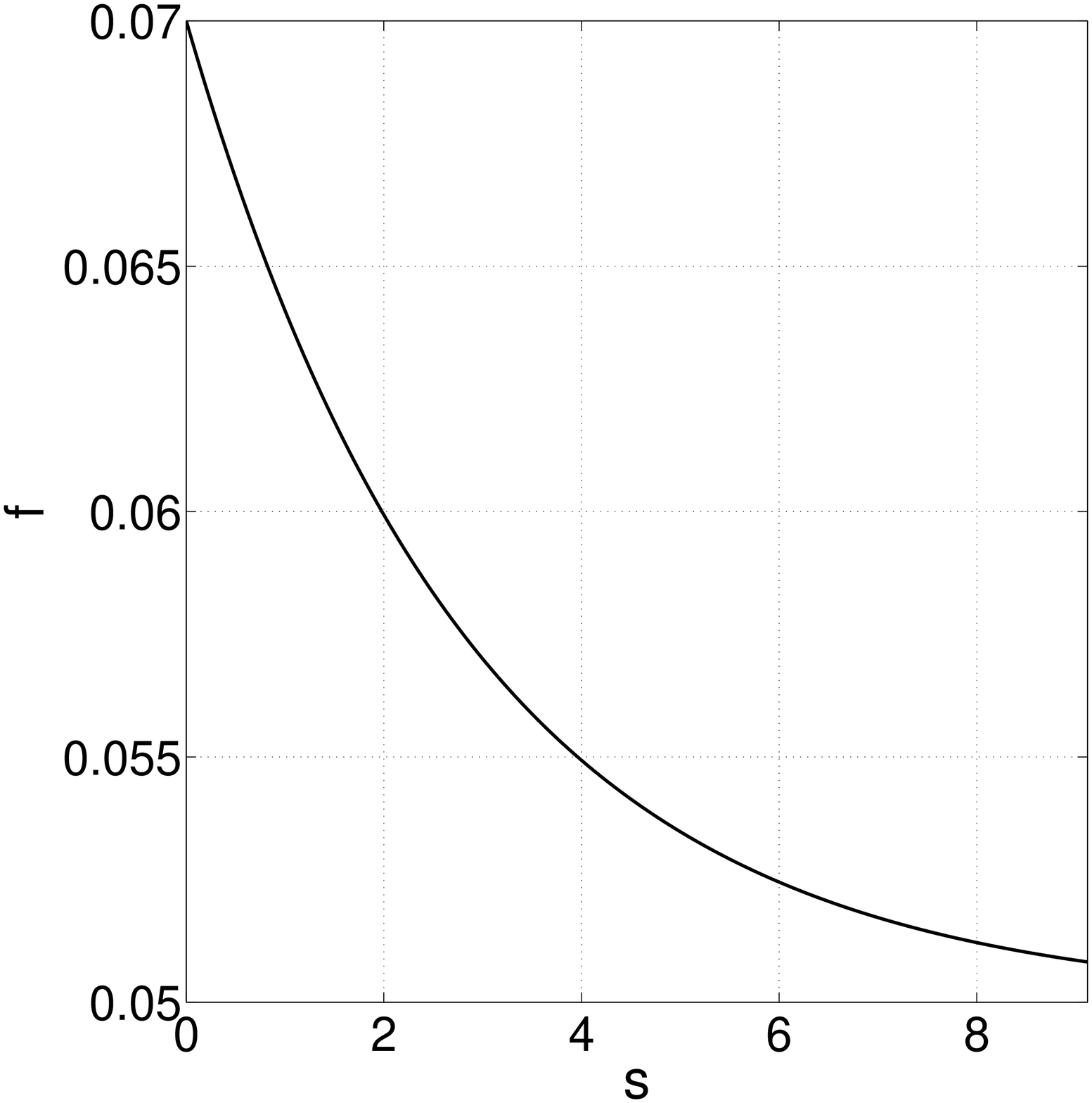}
\end{minipage}
\hspace{-5mm}
\begin{minipage}{0.5\textwidth}\centering
   \psfrag{d}[m][][1][0]{ $d/\unit{\upmu m}$ }
   \psfrag{s}[m][][1][0]{ $s$ }
   \includegraphics[width = 0.7\textwidth]{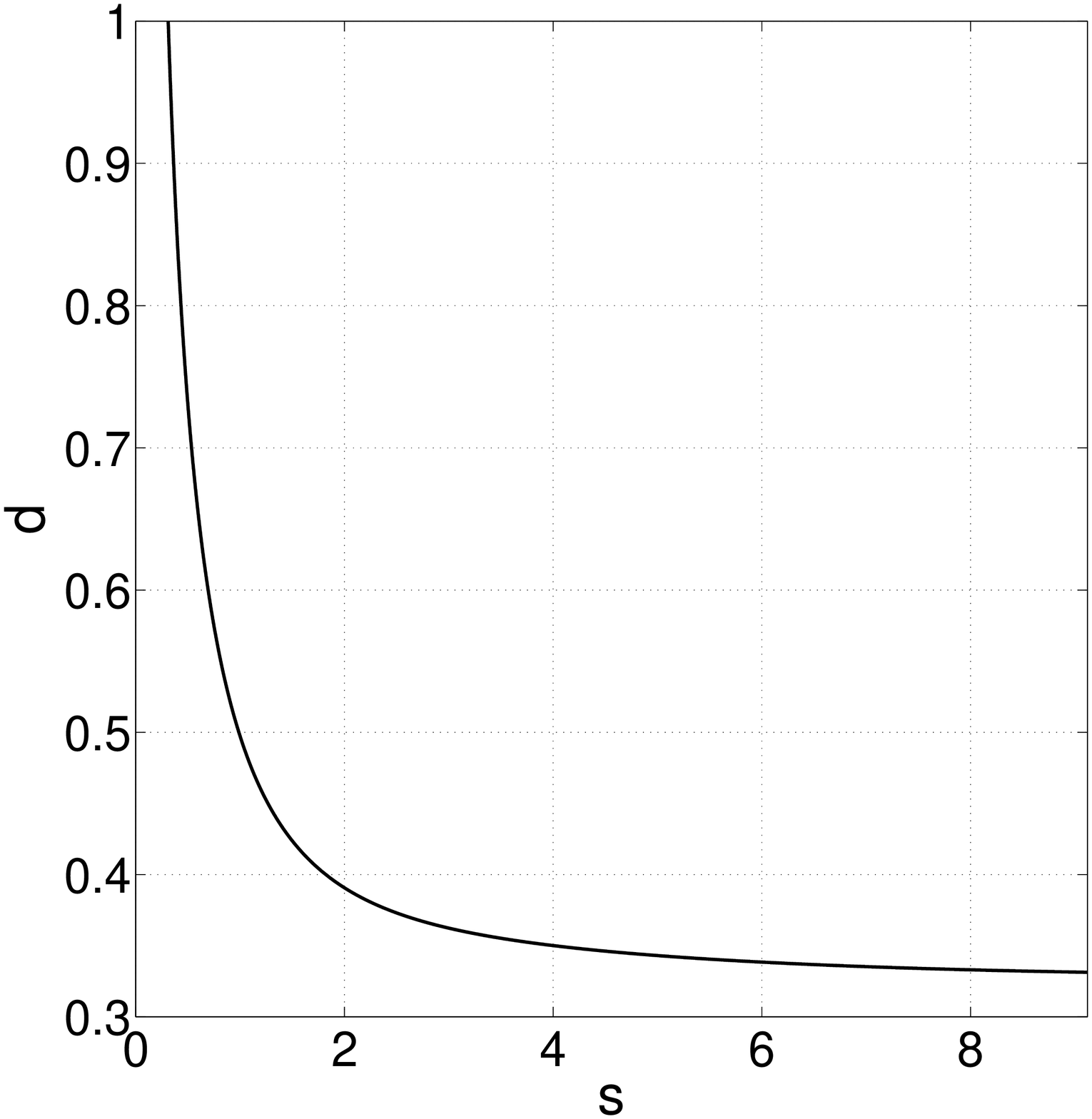}
\end{minipage}
\caption[]{Evolution of the cell wall volume fraction $f$ (left) and size $d$ (right) for ECAP Route $A$ corresponding to Fig. \ref{rwrcrtA}.}\label{fA}
\end{figure}
Hence, the solution of system \eqref{EstrinFin} behaves in a similar way to the solution of the original model \eqref{EstrinOrig1}. In contrast to the numerical results reported by \citet{Mckenzie2007} the dislocation densities $\rho_\w(s)$ and $\rho_\c(s)$ are monotonic functions of $s$ without any change in the sign of curvature for Route $A$.
\par
The load path sensitivity and the resulting non-monotonic course of $f$ for routes $B_c$, $C$ and $E$ enriches the dynamic behaviour of system \eqref{EstrinFin}. Perpetual load path changes lead to periodic behavior (limit cycle in phase space) around a stationary point. As intended by the modeling in Section \ref{sec_FluxWC} the amplitudes of the oscillation of the microstructural state variables $\rho_\w(s), \rho_\c(s), f(s)$ depend on the intensity and duration of the load path change.
\par
Let us confront the numerical results with some more experimental findings. \citet{Viatkina2003} and \citet{Yalcinkaya2009} actually distinguish two scenarios of load path changes with different characteristics. In the case of complete load reversal ($\varTheta=\pi$, proportional non-monotonic loading), a \emph{disruption} of the cells is reported \citep{Viatkina2003}. There, the wall volume fraction $f$ does not change, but the cell size $d$ increases. In the case of non-proportional loading with $\varTheta=\pi/2$ an increase of $f$ but no significant change of $d$ is observed.\vspace{-1mm}
\par
The model proposed in this paper cannot distinguish between these two idealized cases, but it can describe them \emph{by trend}. The response for $\varTheta=\pi$ (cf. Route $C$) is less intense than for  $\varTheta=\pi/2$ (cf. Route $B_c$). Thus, $f$ increases less in the case of complete load reversal whereas $d$ still increases. Moreover, the model offers the advantage that the right hand side of \eqref{EstrinFin} \emph{continuously} depends on load path changes characterized by well-defined tensor invariants.\par
The curves of the predicted cell size show the following result: the less intense the load path changes in the loading history the smaller the achievable cell size. Because the cell size reduction is a precursor of the subgrain structure, this means that Route $A$ would produce cristallites of least size. However, due to the choice of the material parameters implemented in this study, the simulation results can only serve as an initial qualitative estimation of these effects. \par
Summarizing this subsection we conclude that load path changes accompanying procedures of severe plastic deformation could essentially influence the resulting nanocrystalline microstructure. The obtained evolution of dislocation populations differs signficantly depending on whether ECAP route $A$, $C$, $B_c$ or $E$ is applied.

\section{Conclusions and Discussion} 

In this paper, an existing dislocation cell model \citep[cf.][]{Estrin1998,Mckenzie2007} with two dislocation populations was carefully analyzed, refined and extended. For the refined model, microstructural parameters were identified by minimizing a least squares error function. In a further step, the evolution of the microstructural state variables was made sensitive to load path changes. In particular, experimental knowledge about the dissolution and disruption of dislocation cells was incorporated. Tensor-valued internal variables were utilized to capture load path changes. In order to provide these quantities, the microstructural model was coupled with a phenomenological model of finite-strain viscoplasticity. Thereby, special emphasis was put on defining a \emph{physically interpretable} measure of local load path changes. The resulting set of equations was hence called ``load path sensitive two-population dislocation cell model''. The proposed approach was then exemplified by the simulation of various load cases as typical for ECAP routes.\par
The numerical results can now be summarized. It was found that the smallest cell size can be achieved in the case of pure monotonic deformation according to ECAP route $A$. Due to the temporary cell dissolution shortly after load path changes, ECAP routes $C$, $B_c$ or $E$ could slowdown the shrinkage of the cell size. It is worth noting that even though the transient phases of cell dissolution seem to be short related to the entire process, they could signficantly affect the final cell size and subgrain size respectively.\par
In the present study the proposed ``load path sensitive two-population dislocation cell model'' was considered to be a microscopic model. The coupling to the macroscopic continuum model of viscoplasticity was performed in order to account for an aspect of the evolving microstructure. Both models cannot explicitly consider the discrete nature of dislocation motion. Noting that only the evolution of dislocation \emph{densities} is considered reveals some limitations of the proposed approach. The collective behavior of dislocations is much more complicated under real conditions. Hence, other effects could also strongly influence the pattern formation and cell size evolution during non-proportional loading. However, the development of a practicable approach based on fundamental principles of dislocation kinetics which allows for a description of the formation \emph{and} dissolution of dislocation patterns is an open problem. Thus, the current study presents a reasonable idealization of the aforementioned phenomena as it is able to model large systems in long time frames.\par
In order to keep the model practically applicable, simplified phenomenological assumptions were made. It should be noted that the numerical predictions of the evolution of dislocation densities reflect the real behaviour in a qualitative rather than in a quantitative way. Nevertheless, the obtained results seem reasonable and might provide some insights into the dislocation cell evolution under temporary cell dissolution. Still, it is an open question whether the predicted oscillation of the cell size can be observed experimentally. Adequate experimental investigations should include the measurement of cell size evolution shortly after the load path change, which seems to be a challenging task. Furthermore, to \emph{reliably} identify micromechanical parameters, corresponding experiments must be designed in such a way that each parameter which appears in \eqref{EstrinFin} takes effect. Ideally, a set of microstructural measurements at varying temperatures, hydrostatic pressures and plastic strain rates from a procedure of severe plastic deformation is required. Furthermore, it seems promising to investigate the impact of non-proportional loading on dislocation patterns more in detail. In this way, more information about the proposed intensity function $\Inten(\varTheta)$ could be gathered.\par
The performed numerical simulations reveal an interesting result: under the assumptions that the volume fraction of the cell walls increases and a reverse dislocation flux from the wall to the cell interior is initiated shortly after the load path change, a drop of the total dislocation density is predicted. This agrees well with experimental findings \citep{Hasegawa1975}. Moreover, this type of behavior was not explicitly introduced into the model, but results naturally from the aforementioned assumptions.\par
In the current study, there is only an effect from the macroscopic on the microscopic state variables, but no dependence of the macro model on the micro model. This feedback can be easily introduced by making the phenomenological equations \eqref{EvuOrig} - \eqref{EvuOrig2} dependent on the dislocation population densities or the cell wall volume fraction. Additionally, the temperature input to the load path sensitive two-population model \eqref{EstrinFin} can be provided by the thermoviscoplastic model of \citet{Shutov2011}. With the help of the coupled model, the solution of initial boundary value problems will be possible. In particular, more realistic simulations of severe plastic deformation procedures related to heat generation will become possible.

\section*{Acknowledgement}

This research was supported by German Science Foundation (DFG) within SFB 692. The authors are grateful to Prof. M.F.-X. Wagner, Prof. S. Groh and Prof. T. Halle for fruitful discussions.

\bibliographystyle{elsarticle-harv}
\bibliography{bibliography_db}







\end{document}